\documentclass[12pt]{article}

\usepackage{amssymb,amsmath,amsfonts,eurosym,geometry,ulem,graphicx,caption,color,setspace,sectsty,comment,footmisc,caption,natbib,pdflscape,array, graphicx, float, booktabs, appendix, tabularx, mathtools, threeparttable, rotating, lscape, enumerate, bbm, subfig,multirow,enumitem}

\usepackage{hyperref}
\hypersetup{
colorlinks=true,
citecolor=blue,
linkcolor=blue,
filecolor=blue,      
urlcolor=blue,
}

\DeclareMathOperator*{\argmin}{arg\,min}
\newcommand{\bX}{\mathbf{X}}

\newcolumntype{L}[1]{>{\raggedright\let\newline\\arraybackslash\hspace{0pt}}m{#1}}
\newcolumntype{C}[1]{>{\centering\let\newline\\arraybackslash\hspace{0pt}}m{#1}}
\newcolumntype{R}[1]{>{\raggedleft\let\newline\\arraybackslash\hspace{0pt}}m{#1}}

\begin{document}

\begin{titlepage}
\title{Predicting Exporters with Machine Learning\footnotetext{\scriptsize{We want to thank Tommaso Aquilante, Gabor Bekes, Kristina Bluwstein, Falco Bargagli Stoffi, Mahdi Ghodsi, Andreas Joseph, Massimo Riccaboni, Michele Ruta, Gianluca Santoni, Beata Smarzynska Javorcik, Maurizio Zanardi for valuable comments. We want also to acknowledge valuable suggestions by participants to the  annual meeting of the European Economic Association 2022, the seminar series jointly organized by FIW and WiiW in Vienna, and to the European Trade Study Group 2021 in Ghent. Armando Rungi claims financial support from Artes 4.0 - Industry 4.0 Competence Center on Enabling Digital Technologies and Systems.}}} 
\author{Francesca Micocci\thanks{\scriptsize{Mail to: \href{mailto://francesca.micocci@imtlucca.it}{\color{blue}francesca.micocci@imtlucca.it}. Laboratory for the Analysis of Complex Economic Systems, IMT School for Advanced Studies, piazza San Francesco 19 - 55100 Lucca, Italy.} } \and Armando Rungi\thanks{\scriptsize Mail to: \href{mailto://armando.rungi@imtlucca.it}{\color{blue}armando.rungi@imtlucca.it}. Laboratory for the Analysis of Complex Economic Systems, IMT School for Advanced Studies, piazza San Francesco 19 - 55100 Lucca, Italy.}}

\date{\small{This version: September 2022} }
\maketitle
\vspace{0.01in}
\begin{abstract}
\footnotesize
\noindent In this contribution, we exploit machine learning techniques to evaluate whether and how close firms are to become successful exporters. First, we train and test various algorithms using financial information on both exporters and non-exporters in France in 2010-2018. Thus, we show that we are able to predict the distance of non-exporters from export status. In particular, we find that a Bayesian Additive Regression Tree with Missingness In Attributes (BART-MIA) performs better than other techniques with an accuracy of up to $0.90$. Predictions are robust to changes in definitions of exporters and in the presence of discontinuous exporting activity. Eventually, we discuss how our exporting scores can be helpful for trade promotion, trade credit, and assessing aggregate trade potential. For example, back-of-the-envelope estimates show that a representative firm with just below-average exporting scores needs up to $44\%$ more cash resources and up to $2.5$ times more capital to get to foreign markets.

\vspace{30pt}
\bigskip
\end{abstract}
\setcounter{page}{0}
\thispagestyle{empty}
\end{titlepage}
\pagebreak \newpage

\onehalfspacing

\section{Introduction}\label{sec: intro}

Building trade capacity is a purpose of many international and national agencies. The World Trade Organization provides special support programs for developing countries to better integrate into the multilateral trading system. On the other hand, many developing and developed economies prefer to establish their facilitative agencies to provide firms with information, technical advice, marketing services, and policy advocacy about access to foreign markets.\\

The general idea is that there are opportunities for gains from trade, yet not all firms have the same ability to sell their goods and services abroad. Exporting activity entails beach-head costs when handling different regulatory environments, meeting different consumer tastes, and establishing marketing and logistics channels. Only some more productive firms may be able to self-select into exporting status. In contrast, other companies may not have the necessary skills or resources to propose in foreign markets\footnote{For a review of the arguments according to which only the most efficient firms can self-select into an export status and the consequences on the sources of gains from trade, see among others \cite{bernard1999exceptional, Bernardetal2012, Melitz&Redding2014, Hottman&Redding&Weinstein}}. Hence, the necessity to resort to trade promotion programs to fill the gap and help firms build trade capacity to take advantage of open markets. Eventually, openness to trade is a determinant of economic growth insofar as it allows exploiting differential comparative advantages and economies of scale. Companies can benefit while tapping into foreign technology and raising aggregate productivity in the home countries \footnote{Seminal works identify macroeconomic linkages between trade openness, technological progress, and economic growth. See \cite{Grossman&Helpman1990}, \cite{RiveraBatizRomer1991}, \cite{Romer1994}, \cite{BarroSalaimartin1997}.}.\\

Against the previous background, our simple intuition is to adopt machine learning techniques to evaluate how far a company is from reaching an export status based on the assumption that firms' accounts convey non-trivial information on firm-level trade capacity. In other words, we propose to train an algorithm on in-sample financial statements to predict out-of-sample firms' ability to start exporting. Our intuition follows what financial institutions make to predict credit risk, for example, in the case of traditional Altman's Z-scores \citep{altman1968financial} or Merton's Distance-to-Default \citep{merton1974pricing}. Different from credit risk literature, our problem is not to check if a company is proximate to being bankrupt. On the contrary, our challenge is to measure how far a company is from being healthy enough to start and propose on foreign markets.\\ 

We begin by training different machine learning techniques on a sample of 57,016 manufacturing firms in France, which may have exported or not in 2010-2018. Following statistical standards, we randomly partition the initial sample in an 80-20 proportion to separate it into a training and a testing set. Therefore, we train different models armed with a battery of $52$ predictors that we believe may contain non-trivial information on exporting abilities. Then we use the trained models to obtain distributions of out-of-sample predictions that can be useful to assess a company's distance from exporting capability. In simple terms, the exporting score summarizes how much a non-exporter looks like an exporter.

Crucially, we find that our procedure correctly separates exporters from non-exporters with an accuracy of up to $90\%$. The latter is a figure we obtain from a horse race among different algorithms. We find that a Bayesian Additive Regression Tree with Missingness not at Random (BART-MIA) \citep{kapelner2015prediction} is the procedure that provides the most robust predictions. The BART-MIA is a regression tree with a Bayesian component for regularization through a prior specification that allows flexibility in fitting various regression models while avoiding strong parametric assumptions \citep{HillLineroMurray}. What makes BART-MIA especially useful for our case is the possibility of exploiting additional predictive power from non-random missing values on predictors. The latter is a feature that is especially useful in catching business dynamics when coverage of financial accounts is likely to be correlated with other dimensions, e.g., firms' size or productivity, which, in turn, can correlate with firms' export status. In our case, we assess that considering non-random missing values helps us increase prediction accuracy by about $14.4\%$. 
Eventually, we ensure that prediction accuracies are robust to different definitions of exporters and to the presence of discontinuous exporting activity \citep{geishecker2019one, BekesMurakozy}. The last check is especially relevant in the case of smaller exporters, or when exporters specialize in manufacturing capital goods, whose relationships with customers entail several breaks in the time series. 

Our framework is also robust to different cross-validation strategies since we obtain similar performance by randomly picking training and testing subsets in different ways, albeit from a unique sample. Finally, we test that reducing the set of predictors brings lower levels of accuracy after we perform a Least Absolute Shrinkage and Selection Operator (LASSO) for dimensionality reduction \citep{belloni2013least, belloni2014high, belloni2016inference, robustlasso}.

After assessing which tool is better at predicting exporters, we delve into the prediction power of single predictors, i.e., how much they contribute to get good predictions. The practical utility of this exercise is to show that there may be indeed some dimensions of the firms' economic activity that correlate relatively more with their trade potential. Thus, following \citet{chipman2010bart}, we implement a procedure to derive \textit{Variable Inclusion Proportions} (VIPs), which can be interpreted as posterior probabilities \citep{bleich2014variable}. Crucially, we discuss how VIPs have a relevant internal validity, since they catch predictive power within the given testing vs. training sets. Yet, we may not attribute them any external validity because predictors can change their power in different contexts. Indeed, we discuss how such changes in different contexts and sub-populations could actually be informative of the changing resilience of firms and from where it comes.
For example, in the French case we study,the difference we observe in the model's selection of influential predictors between \^{I}le-de-France and the rest of France  suggests there are geographic-specific firms' dynamics. The same predictors may or may not play a major role on the probability of exporting, depending on the the specific technological characteristics of the production environment.


Final sections discuss how we see exporting scores applied in practice. We suggest looking at baseline predictions to derive a probabilistic exporting score to a firm, i.e., a score summarising how similar a non-exporter is to benchmark exporters on a scale from $0$ to $1$. We argue that exporting scores could be helpful for trade promotion or trade finance programmes. After aggregation, we show how they can represent an additional tool to describe the trade competitiveness of regions or industries.

Finally, to briefly illustrate the practical utility of exporting scores, we classify firms into risk categories and provide simple back-of-the-envelope estimates of how much cash resources and capital expenses they would need to reach export status. We find that increasing cash and capital is required to reduce the distance from export status. For example, in the case of medium-risk firms, i.e., firms that have just below $50\%$ probability of exporting, we show a need for up to $44\%$ more cash resources and up to $246\%$ more capital expenses to reach full export status.

The remainder of the paper is organized as follows. In Section \ref{sec: literature} we relate to previous literature. We introduce data and sample coverage in Section \ref{sec: data}, whereas Section \ref{sec: strategy} discusses the empirical strategy. Results are commented in Section \ref{sec: results}, while robustness checks are discussed in Section \ref{sec: robustness}. A specific Section \ref{sec: discontinuous} tests for the sensitivity of predictions to the phenomenon of temporary trade, while the practical use of exporting scores is offered in Section \ref{sec: scoring}. Section \ref{sec: conclusions} concludes.

\section{Related literature}\label{sec: literature}

Most countries worldwide implement trade promotion programmes that envisage the expenditure of substantial amounts of public funds. Thus, it is hardly surprising that there have been concerns about the efficacy and effectiveness of those support programs. Interestingly, \cite{Volpe&Carballo2008} show how export promotion actions are usefully associated with increased exports by already trading firms and traded products, i.e., the intensive margin. In terms of extensive margins, i.e., the increase of firms and products crossing national borders, \cite{Volpe&Estevadeordal&Gallo&Luna2010} show that an influential role is often played by the establishment of diplomatic representations, especially in the case of producers of homogeneous goods. In general, activating new trading relationships may require various services bundled into more complex export promotion programmes \citep{Volpe&Carballo2010Services}. Eventually, a majority of studies investigate how effective a policy is on the \textit{ex-post} companies' exporting performances while controlling for cherry-picking \citet{Volpe&Carballo2010}. In general, \cite{VanBiesebroecketal2016} demonstrate how trade promotion programmes have been a vital tool to overcome economic crises, as in the case of recovery after the global recession in 2008-2009.

In this context, our contribution focuses explicitly on the possibility of increasing the trade extensive margin proposing a measure of the ability of non-exporters to start exporting. From this perspective, what we propose is a pure prediction exercise based on the intuition that exporters are statistically different from non-exporters. In this sense, we rely on a two-decades-long strand of research that has established a connection between firms' heterogeneity and trading status \citep{bernard1999exceptional, melitz2003impact, Melitz&Ottaviano2008, Bernardetal2012, Melitz&Redding2014, Hottman&Redding&Weinstein}. Our intuition is that a prediction of export status is possible only because we have prior knowledge that exporters do have different cost structures than non-exporters. After all, they have to sustain the fixed costs to gain access to foreign markets, where regulations and consumer tastes can differ much from home and where shipping is costly. Thus, we demonstrate that starting from a comprehensive battery of economic and financial predictors allows indeed separating exporters from non-exporters with a relatively high prediction accuracy, up to $90\%$. 

Please note that ours is not a classic policy evaluation exercise nor a structural model to understand the determinants of export status. We do not want to assess whether any specific policy design works to support would-be exporters. Ours is a simple scoring exercise in the fashion of what one can find in previous literature about credit scoring, where there is a long tradition to try and spot firms in financial distress based on the disclosure of financial accounts. See seminal attempts with Z-scores by \cite{altman1968financial, altman2000predicting}, and Distance-to-Default by \cite{merton1974pricing}, where some specific threshold is set as a rule of thumb to say whether a firm is financially sound and worthy of credit. Nowadays, most financial institutions adopt predictive models to evaluate credit risk, including machine learning \citep{Uddin2021}. See the exercises on firm-level correlations to spot investment-to-cash-flow sensitivities and assess time-varying financial constraints \citep{fazzarihubbardpetersen, Almeida&Campello&Weisbach2004, Chen2012}. 

The additional difficulty in our exercise is that we want to score success, i.e., the ability of a firm to outreach across national borders. In contrast, credit risk analyses take as reference previous firms' failures, i.e., their distance-to-default. Yet, we argue, the intuition is the same: to get as a benchmark firms that realized an outcome, in our case an export status, and thus measure how far we are from that outcome. 

Eventually, we can also relate to literature on trade finance. We know very well that routine access to trade credit is needed to outlive foreign markets, and well-functioning ﬁnancial markets are crucial to export performance \citep{Manova2012}. Eventually, external finance helps firms to gain and keep access to foreign markets despite the high beach-head costs, especially in the case of smaller producers who have a reduced ability to provide collateral to financial institutions \citep{Chor&Manova2012}. In this context, we believe exporting scores are potentially valuable to better target financial institutions' credit policies in a familiar way, e.g., by considering credit risk classes. 

To better grasp our previous intuitions, we propose a very simple back-of-the-envelope exercise that estimates, \textit{ceteris-paribus}, how much cash resources and capital expenses firms need to switch across low, medium and high-risk classes.

Moreover, from a macroeconomic viewpoint, one can use firms' scoring as yet another indicator of the competitiveness of an economy (or lack thereof). Inspired by so-called growth diagnostics, international and national statistics offices have developed frameworks for assessing the potential of countries, regions, and industries to compete ion international markets. See, for example, the work by the World Bank on measuring trade competitiveness \citep{Reis&Wagle&Farole, Gaulieretal2013}. In the case of French manufacturing, we show how potential exporters are unevenly distributed across industries and regions. We believe there is no reason why an indicator like ours about the potential of extensive margins should not find room in a standard trade diagnostic kit. 

Finally, we want to remark how ours is one of the first attempts to exploit statistical learning techniques in international economics. As far as we know, there are only a few notable efforts in progress, including \citet{Gopinathetal2020} and \citet{Breinlichetal2021}. Yet, we firmly believe that statistical learning exercises have a great potential and should find their way in a field like international economics, where one often needs to extract valuable information from big and complex datasets, which can be dealt with by a combination of both predictive tasks and standard causal inference exercises \citep{athey2018impact, mullainathan2017machine}.

\section{Data} \label{sec: data}

We source firm-level information from ORBIS\footnote{The ORBIS database has become a standard source for global firm-level financial accounts. For a previous usage of this database, among others, see \cite{gopinath2017capital}, \cite{cravinolevchenko}, \cite{DelPrete&Rungi2017}, and \cite{RungiDelPrete2018}. It complements financial accounts with other information from different sources on ownership, corporate governance, and intellectual property rights, which we also use for predictions in the following analyses.} compiled by the Bureau Van Dijk. Notably, France is a much-explored case study of firm-level trade data that allows us to confront previous literature. See among others \cite{CrozetHeadMayer2011} and \citet{FontagneSecchiTomasi}. Our main outcome of interest is the export status of a firm that we derive from information on export revenues \footnote{Interestingly enough, French firms must report the amount of revenues from exports separately, as from the subsequently amended \textit{Règlement n. 99-03 du Comité de la réglementation comptable.}}. \textit{Prima facie}, we will consider a firm as an exporter if it reports positive export revenues. Then, in Sections \ref{sec: robustness} and \ref{sec: discontinuous}, we will challenge our baseline definition to comply with the phenomenon of temporary trade \citep{BekesMurakozy}, when it is optimal for firms to export every once in a while. As for firm-level predictors of exporting status, we employ a battery of $52$ indicators elaborated on original financial accounts that we use to train our models. Further details on our choice are discussed in Section \ref{sec: predictors}, while we include the list of predictors with a complete description in the Data Appendix.

\begin{figure} [H]
\begin{center}
\caption{Sample coverage: exporters by region}
\includegraphics[width=0.9\textwidth]{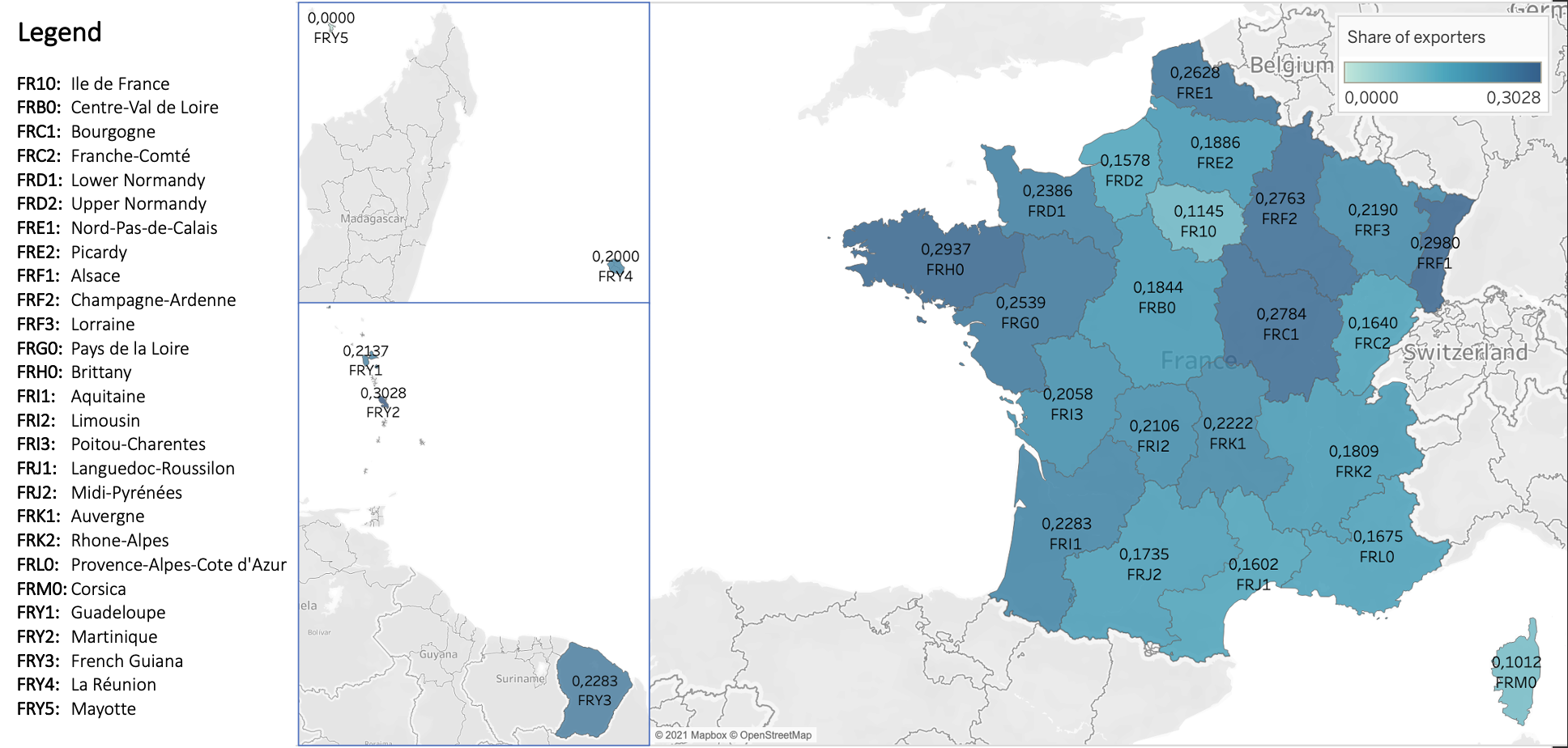}
\label{fig: geo}
\end{center}
\begin{tablenotes}
\item
\footnotesize
\centering
Note: Unitary shares indicate exporters on total firms in NUTS 2-digit regions.
\end{tablenotes}
\end{figure}

To grasp the coverage of our sample, we draw Figure \ref{fig: geo} and Table \ref{tab:industry coverage}. Figure \ref{fig: geo} shows how relevant exporters are in every NUTS-2 region in France, as from our sample. Table \ref{tab:industry coverage} compares sample industry coverage with the one provided by Eurostat census in 2018. We do find that we have a fair coverage by 2-digit industries since the correlation by industry shares is about $0.90$. Yet, our sample covers $32.6\%$ of firms' population that, however, represents about $75\%$ of total operating revenues in France according to Eurostat business demographics. As largely expected, we cannot retrieve financial accounts of smaller firms, because they are not required to comply with accounting regulations in the same way as medium and larger ones. See also a comparison by class categories with Eurostat in Appendix Table \ref{tab: coverage size_industry}. In the following paragraphs, we will show how our baseline analysis can handle non-random missing values in financial information.

\begin{table}[H]
\centering
\caption{Sample coverage by industry}
\label{tab:industry coverage}
\resizebox{0.92\textwidth}{!}{%
\begin{tabular}{lccccccccc}
\hline
 & & \multicolumn{4}{c}{Sample}  &\multicolumn{4}{c}{Population}  \\
NACE rev. 2 & code & non-exporters & exporters & total & (\%) & non-exporters & exporters & total & (\%) \\
\hline
       Food products&10&1,3057&1,429&14,486&0.254&49,153&2,135&51,288&0.293\\
       Beverages&11&1,176&395&1,571&0.028&3,028&825&3,853&0.022\\
       Textiles&13&919&389&1,308&0.023&4,278&798&5,076&0.029\\
       Wearing apparel&14&1,060&336&1,396&0.024&8,813&881&9,694&0.055\\
       Leather and related products&15&374&142&516&0.009&2,930&313&3,243&0.019\\
       Wood and products of wood and cork&16&2,203&509&2,712&0.048&8,920&1,036&9,956&0.057\\
       Paper and paper products&17&455&362&817&0.014&823&469&1,292&0.007\\
       Printing and reproduction of recorded media&18&2,995&584&3,579&0.063&14,347&969&15,316&0.088\\
       Coke and refined petroleum&19&17&14&31&0.001&-&-&25&0.0001\\Chemicals and chemical products&20&958&705&1,663&0.029&1,388&1,127&2,515&0.014\\
       Pharmaceutical products&21&151&148&299&0.005&93&159&252&0.001\\
       Rubber and plastic products&22&1,436&931&2,367&0.042&1,780&1,425&3,205&0.018\\
       Other non-metallic products&23&1,929&393&2,322&0.041&7,026&777&7,803&0.045\\
       Basic metals&24&354&267&621&0.011&295&304&599&0.003\\
       Fabricated metal prod., except machinery and equipment&25&8,135&2,540&10,675&0.187&14,557&3,903&18,460&0.106\\
       Computer, electronic and optical products&26&965&605&1,570&0.028&1,304&991&2,295&0.013\\
       Electrical equipment&27&789&495&1,284&0.023&1,321&727&2,048&0.012\\
       Machinery and equipment&28&1,938&1,194&3,132&0.055&2,567&1,967&4,534&0.026\\
       Motor vehicle, trailers and semi-trailers&29&748&424&1,172&0.021&1,119&516&1,635&0.009\\
       Other transport equipment&30&330&186&516&0.009&847&260&1,107&0.006\\
       Furniture&31&1,416&249&1,665&0.029&8,758&598&9,356&0.053\\
       Other manufacturing&32&2,796&518&3,314&0.058&19,960&1,378&21,338&0.122\\
         \hline
        Total &  & 44,201 & 12,815 & 57,016& 1.00& 153,307 & 21,558 & 174,890 & 1.00\\
        \hline
\end{tabular}}
\begin{tablenotes}
\footnotesize
\singlespacing
\item Note: French manufacturing firms are sourced from Orbis, by Bureau Van Dijk. On columns 3 and 4, we separate exporters and non-exporters in our sample. On column 5 we report the total number of manufacturing firms by NACE rev.2. On columns 7-9 a comparison with Eurostat census.  When we look at shares on columns 6 and 10, we find our sample is well balanced by industry if compared with the population.
\end{tablenotes}
\end{table}

\section{The empirical strategy}\label{sec: strategy}
\onehalfspacing
Our main intuition is that we can predict out-of-sample exporting capability based on the in-sample experience of both exporters and non-exporters. The first step is to find the best algorithm that is able to separate exporters and non-exporters after conditioning on financial information. Our prior is that exporters and non-exporters are statistically different, as acknowledged by previous literature reported in Section \ref{sec: literature}. Thus, once we assess the method that assures the best predictive accuracy with the minimum numbers of false positives and false negatives (see Section \ref{sec: findings}), we can test out-of-sample and use the distribution of predictions to assign each firm an exporting score that is bounded, by construction, in an interval from $0$ to $1$. The higher the score, the better the chances a firm is able to make it on foreign markets. 

\begin{figure} [H]
\begin{center}
\caption{Visual intuition of an exporting score.}
\includegraphics[scale=0.55]{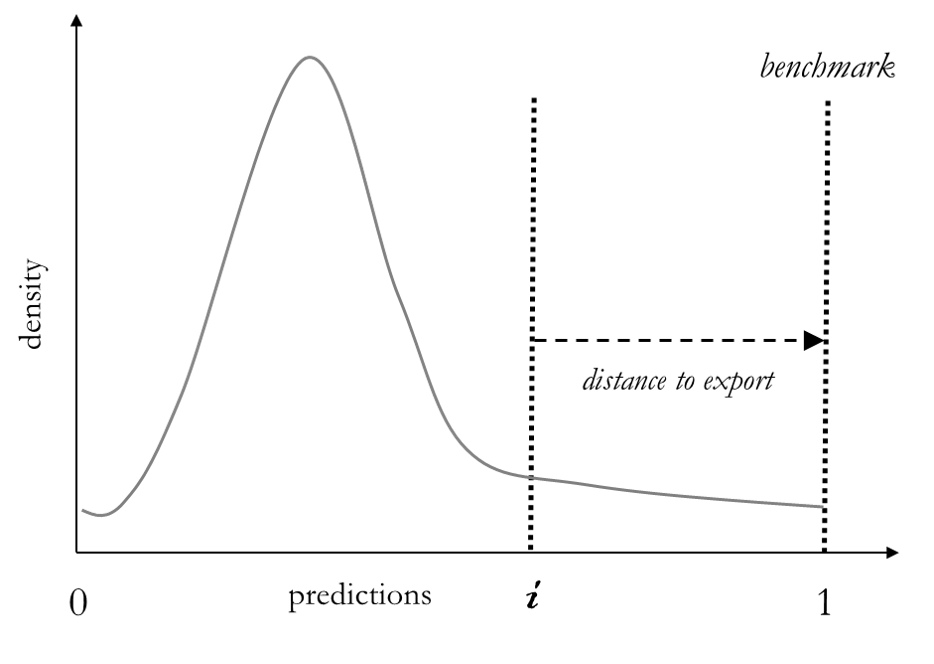}
\label{fig: intuition}
\end{center}
\begin{tablenotes}
\item
\footnotesize

Note: We represent a fictional distribution of predictions of exporting status that is by definition bounded in an interval $[0, 1]$. Along the distribution, we could spot an $i$-th non-exporting firm. We reasonably assume that actual exporters locate at the end of the right tail. By definition, non-exporters are less and less likely to start exporting at an increasing distance from predicted exporters.  
\end{tablenotes}
\end{figure}

In Figure \ref{fig: intuition}, we report a visual fictional representation of our intuition. Assuming that we did a good job in training and that prediction accuracy is acceptable, we can reasonably test on new firms and locate actual exporters at the end of the right tail of the distribution of exporting predictions. Thus, any $i$th non-exporting firm located on the left of predicted exporters will come with a positive distance, which will convey non-trivial information on how viable that firm is to start exporting. In other words, we take as a reference point the export status at $1$ and, thus, we check how far a company is from that reference point.\\ 

Eventually, in Section \ref{sec: coefficients} we provide a framework for the interpretability of predictors by catching the influence of each of them in getting the exporting scores. That is, we are able to sum up how important one predictor is with respect to the entire set in any out-of-sample exercise we may run. Obviously, given the predictive nature of our analyses, we won't be able to attach any causal interpretation to our exercise. For our purpose, we will make use of \textit{Variable Inclusion Proportions}, i.e., the proportion of times a predictor is selected as splitting rule for the construction of the random trees. The construction and interpretation of VIP are discussed in section \ref{sec: coefficients}. Notably, selected predictors are contingent on the trained sample, i.e., their role won't have any external validity. Yet, we argue, identifying the drivers of the model performance helps further comment on the nature of exporting scores.

\subsection{Methods}\label{sec: methods}
To get our best predictions, we train and compare different statistical learning techniques. Thus, we can make use of the generic predictive model for firms' export status in the form:

\begin{equation}
\label{eq: predictive_model}
    {f}(\bX_{i})={Pr}(Y_{i}=1 \:|\: \bX_{i}=x)
\end{equation}

where $Y_{i}$ is the binary outcome that assumes value $1$ if the $i$th firm is exporting, and zero otherwise. $\bX_{i}$ is a $P$-dimensional matrix that includes a full battery of firm-level predictors, which we discuss in detail in the following Section \ref{sec: predictors}. Please note that, at this stage, we do not consider the time dimension, i.e., we train the predictive model considering the export status of a firm in relation with present predictors. In this baseline model, it is entirely possible that a firm is considered as an exporter in one year and a non-exporter in another year. See Section \ref{sec: discontinuous} where we introduce the time dimension, thus looking at heterogeneous exporting patterns.

The functional form that links predictors to outcomes is \textit{ex-ante} unknown and looked for by the generic supervised machine learning technique. We provide an overview of the different methods we use in Section \ref{sec: methods}. The advantage of any of them is to extract information from many predictors, while catching non-linearities that may be present in the association with export status. Briefly, the generic predictive model has to pick the best in-sample loss-minimizing function in the form:

\begin{equation}\label{eq: generic}
\argmin \sum_{i=1}^{N} L(f(x_i), y_i) \: \: \: \:  over  \: \:  \: \: f(\cdot) \in F \: \: \: \:  \: \: \: \: s. \: t.\: \:  \: \: \: \:  \: \: R\big(f(\cdot)\big) \leq c
\end{equation}

where $F$ is a function class from where to pick the specific function $f(\cdot)$. Importantly, $R\big(f(\cdot)\big)$ is the generic regularizer that summarizes the complexity of $f(\cdot)$. The latter is a tool that allows us to solve the common trade-off between an as high as possible in-sample fit and an as high as possible flexibility of the prediction model, able to take on board new out-of-sample information. It is the solution to the so-called bias-variance trade-off. The set of regularizers, $R$'s, will change following standards proposed by each method that we will compare in the following paragraphs. Eventually, any method shall minimize the constrained loss function represented in eq. \ref{eq: generic}, while searching for the function that can be better used to process new out-of-sample information.

As a common strategy across our different models, we will pick at random $80\%$ of our French firms to be considered as in-sample information. We will then use it to train the generic statistical learning algorithm. We will keep the remaining $20\%$ as out-of-sample information to predict export status. Hence, we will be able to assess the accuracy of our predictions within the limit of our data sources. As it is standard in similar exercises, we perform a cross-validation check described in Section \ref{sec: robustness}, to verify that a specific segment of the sample does not affect prediction accuracy.

In the following paragraphs, we show how a specific variant of the Bayesian Additive Regression Tree (BART) performs better than others, because it is able to consider the presence of non-random missing values as further predictors for the outcome. The variant we use is the BART with Missingness In Attributes (BART-MIA). For more details, see also \cite{kapelner2015prediction}. For a previous application to firms' dynamics, see \cite{zombie}.

In general, any classification tree $\mathcal{T}$ is built on \textit{if-then} statements that split the training data according to the observed values of predictors, allowing for non-linear relationships between the predictors and the outcomes. Thus, the generic algorithm for the construction of a classification tree, $\mathcal{T}$, is based on a top-down approach that recursively splits the main sample into non-overlapping sub-samples (i.e. the nodes and the leaves). Therefore, the tree is pruned iteratively with the generic regularizer $R$ to improve its predictive ability while avoiding overfitting, in case trees develop along too many layers \footnote{It is beyond the scope of this paper to get into further details of single techniques. For a deeper introduction to statistical learning, we refer to \cite{elements}. }.

As in the baseline version \citep{chipman2010bart}, BART-MIA is a sum-of-trees ensemble, with an estimation approach relying on a fully Bayesian probability model. The algorithm elaborates the ensamble by imposing a set of Bayesian priors that regularize the fit by keeping the individual trees' effects small in an adaptive way. The result is a sum of trees, each of which explains a small and different portion of the predictive function. The BART-MIA variant we adopt can be expressed as:
\begin{equation}\label{eq: tree}  \centering 
    \mathbb{P}(Y=1|\bX) = \Phi\left( \mathcal{T}_1^{\mathcal{M}}(\bX) + ... + \mathcal{T}_q^{\mathcal{M}}(\bX)\right),
\end{equation}
where $\Phi$ denotes the cumulative density function of the standard normal distribution and the $q$ distinct binary trees are denoted by $\mathcal{T}$, each being a single tree coming with an entire structure made of nodes and leaves. The sum-of-trees model serves as an estimate of the conditional probit at $\bX$, which can be easily transformed into a conditional probability estimate of $Y=1$.\footnote{Note that each classification probability $P(Y=1|\bX)$ is obtained as a function of a sum of regression trees, while standard classifier approaches use a majority or an average vote based on an ensemble of classification trees. See, for example, \citet{breiman2001random}.}. The Bayesian component of the BART includes three priors that have demonstrated to use efficiently the data at disposal: 
\begin{enumerate}
    \item the prior on the probability that a node will split at depth $k$ is $\beta(1+k)^{-\eta}$, where $\beta \in (0,1), \eta \in [0, \infty)$, and the hyper-parameters are chosen to be $\eta=2$ and $\beta = 0.95$;
    \item the prior on the probability distribution in the leaves is a normal distribution with zero mean: $\mathcal{N}(0, \sigma^2_q)$, where $\sigma_q = 3/d\sqrt{q}$ and $d=2$;
    \item the prior on the error variance is $\sigma^2 =1$.
\end{enumerate}
 Thus, the regularization parameter $R(\cdot)$ in the general formulation of ML algorithm \ref{eq: generic} corresponds to the priors themselves. Finally, the BART-MIA algorithm employs a Metropolis-within-Gibbs sampler \citep{geman1984stochastic,hastings1970monte} to generate draws from the posterior distribution of $\mathbb{P}(\mathcal{T}_1^\mathcal{M},...,\mathcal{T}_m^\mathcal{M},1 | \Phi(Y))$. \footnote{This passage involves introducing small perturbations to the tree structure: growing a terminal node by adding two child nodes, pruning two child nodes (rendering their parent node terminal), or changing a split rule.} Let us denote with $K$ the size of the sample of the draws $\{p^*_1,\dots, p^*_K\}$ from the posterior distribution. Then, the prediction $p(x)=P(Y=1|\bX)$ at a particular $x$, is
\begin{equation*}
    p^*(x)=\sum_{k=1}^K p_k^*(x)
\end{equation*}

In addition to the Bayesian component, the BART-MIA variant augments the original algorithm by exploiting information on missing values and splitting on \textit{missingness} features that are used as additional predictors in each binary-tree component.

Eventually, the BART-MIA is chosen in the following paragraphs as the baseline method after a comparison with four other alternatives. At first, we compare with a simple logistic regression (LOGIT). The latter is a classical econometric technique for binary outcomes with a specific \textit{ex-ante} assumption on the functional form linking predictors with the outcome. Then, we perform three other methods based on regression trees, namely a Classification and Regression Tree (CART) \citep{friedman1984classification}, a Random Forest (RF) \citep{breiman2001random}, and the original unaugmented BART. CART is the most basic regression tree, while RF is an ensemble method that aggregates different regression trees to get a stronger predictive power, as the BART does, but without a Bayesian framework. Finally, we compare previous regression trees' models with the Least Absolute Shrinkage and Selection Operator (LASSO), in the form:

\begin{eqnarray}\label{eq:LOGITLASSO}
\argmin_{{\beta} \in \mathbb{R}^{p}} \:\: \frac{1}{2N} \sum_{i=1}^N \left(y_{i}(x_{i}^T\beta) - log(1 + e^{(x_{i}^T\beta)})\right)^2 \:\:\:
{\rm subject\,\,to\,\,} \|{\beta}\|_1 \leq k.
\end{eqnarray}

where $y_{i}$ is a binary variable equal to one if a firm $i$ is an exporter and zero otherwise. Any $x_{i}$ is a predictor chosen in $\mathbb{R}^{p}$, whereas $\|{\beta}\|_1=\sum_{j=1}^p |\beta_j|$ and $k > 0$. The constraint $\|{\beta}\|_1 \leq k$ limits the complexity of the model to avoid overfitting, and $k$ is chosen, following \citet{robustlasso}, as the value that maximises the Extended Bayesian Information Criteria \citep{chen2008extended}. To account for the potential presence of heteroskedastic, non-Gaussian and cluster-dependent errors, we adopt the rigorous penalization introduced by \cite{belloni2016inference}. 

\subsection{Predictors} \label{sec: predictors}

To increase models' predictability, we include a full battery of $52$ predictors that we derive from firms' balance sheets and profit and loss accounts. A detailed description is reported in the Data Appendix. Broadly speaking, we choose to include:

\begin{enumerate}
    \item original financial accounts without any elaboration;
    \item financial ratios and other proxy indicators (e.g., productivity, economies of scale, spillovers) that we expect to be correlated with exporting activity;
    \item firms' locations, ownership status, and industry affiliations, which can help in spotting categories of firms at a competitive advantage or disadvantage.
\end{enumerate}

\begin{figure}
    \centering
    \caption{Correlation matrix of predictors}
            \resizebox{0.75\textwidth}{!}{%

    \includegraphics[width=\textwidth]{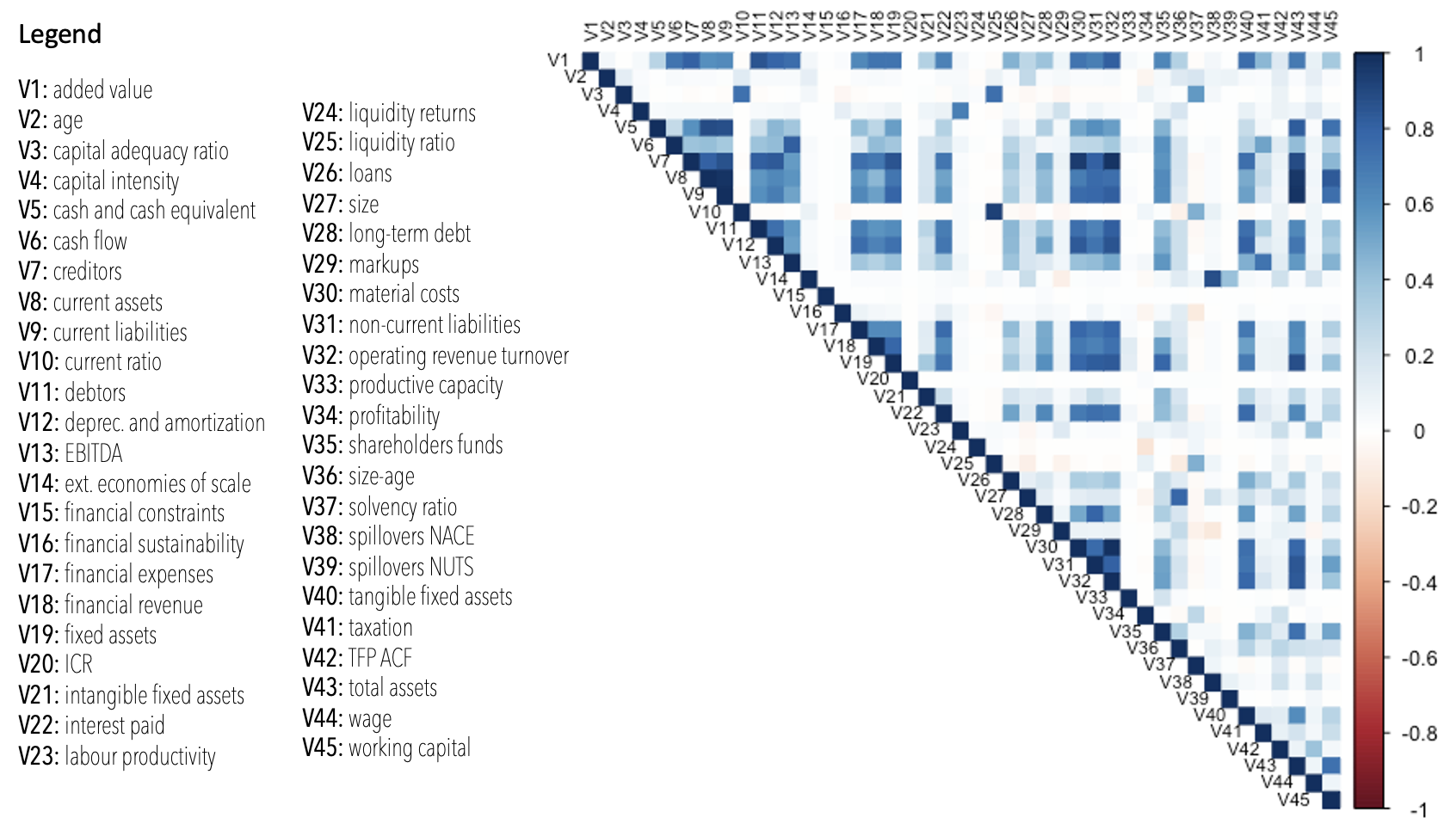}}
    \label{fig: corr_mat_pred}
        \begin{tablenotes}
\footnotesize
\singlespacing
\item Note: We report a correlation matrix of the predictors we use. Non-numeric predictors are excluded here but included in following analyses: NUTS-2 locations, NACE Rev.2 industries, a categorical variable for consolidated accounts, patents' dummy, inward FDI, outward FDI, and corporate control. Positive correlations are reported as shades of blue, while negative correlations are reported as shades of red.   
\end{tablenotes}
\end{figure}

Usefully, in Figure \ref{fig: corr_mat_pred}, we show a correlation matrix including all numeric predictors. Please note how some of them are indeed much cross-correlated with values well above $0.6$. Yet, high correlations are not that relevant to our case since, in a context of pure prediction like ours, we do not (want to) estimate coefficients. At this stage, we also do not need a prior on which financial information conveys the highest predictive power. Hence, we choose not to discriminate among predictors \textit{ex ante}, although we do have information provided by previous literature that some variables  more than others are associated with exporting activity (productivity, firm size, financial constraints, etc.). See also a specific robustness check in Section \ref{sec: robustness}, where we show what happens when we reduce our set of predictors. In another words, we are well aware that our long list of predictors entails a great deal of endogeneity among variables that are otherwise studied in different structural relationships. As we are not interested in obtaining estimates for determinants of trade, such endogeneity is not relevant for our purpose. What we need to do is to minimize the prediction errors given albeit marginally useful observable information. In Section \ref{sec: internal validity}, we further discuss the limits and benefits of a pure predictive exercise when it comes to the interpretability of predictors.

\newpage
\section{Results}\label{sec: results}
\subsection{Models' horse race}\label{sec: findings}

\begin{figure}[!htbp]
  \centering
  \caption{Out-of-sample Goodness-of-Fit}
  \includegraphics[width=0.85\textwidth]{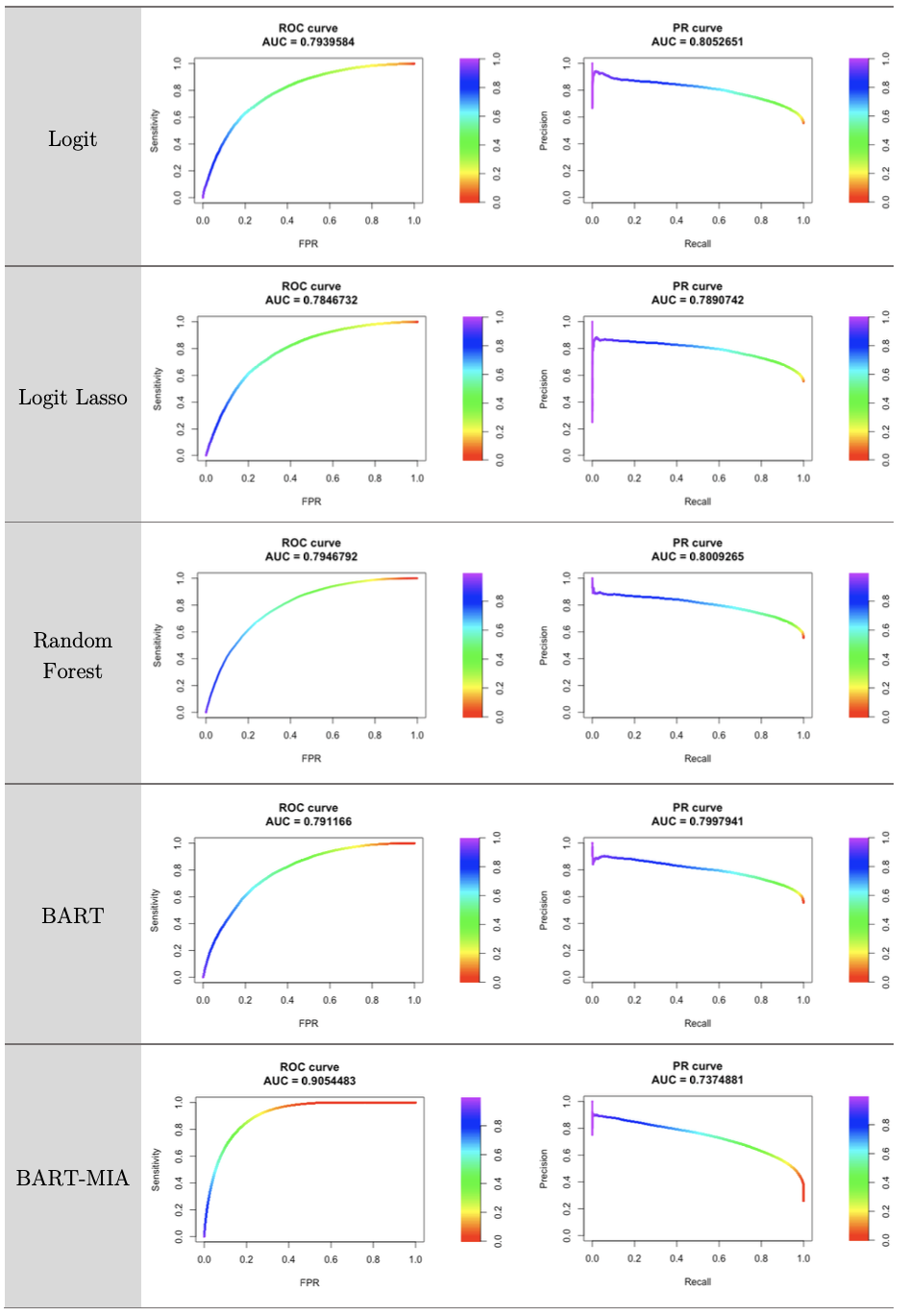}\label{fig:diagnostics}
  \begin{tablenotes}
\footnotesize
\singlespacing
\item Note: We report the ROC Curves and Precision-Recall curves of the models. See Appendix \ref{appendix : c} for the details on the construction of the curves and their interpretation.
\end{tablenotes}
  \end{figure}

In Table \ref{tab:horse_race}, we compare measures of standard prediction accuracy across the methods we test. For details on how these metrics are constructed, please see Appendix C. Briefly, what we can see is that Sensitivity focuses on the ability to predict exporters, i.e., the amount of \textit{true positives}, while Specificity focuses on the ability to predict non-exporters, i.e., the amount of \textit{true negatives}. Balanced Accuracy is an arithmetic mean between Sensitivity and Specificity. Importantly, the ROC curve (receiver operating characteristic curve) evaluates the predictive performance at different classification thresholds, as reported in Figure \ref{fig:diagnostics}, and it is our baseline measure of performance across different models. Finally, Precision-Recall is of help to us in assessing the trade-off between returning accurate results (high precision) vis \'a vis returning a majority of positive results (high recall). 

\begin{table}[H]
    \centering
        \caption{Prediction accuracies}
        \label{tab:horse_race}
        \resizebox{0.8\textwidth}{!}{%
    \begin{tabular}{lcccccr}
    \hline\hline
    &Specificity&Sensitivity&Balanced&ROC&PR& N. obs.\\
    &&&Accuracy&&&\\
    \hline
  LOGIT&0.6642&0.7776&0.7210&0.7940&0.8053&86,754\\
  LOGIT-LASSO&0.6606&0.7722&0.7164&0.7847&0.7891&86,754\\
  CART&0.5700&0.7896&0.6796&-&-&86,754\\
  Random Forest&0.6078&0.8276&0.7178&0.7947&0.8010&86,754\\
  BART&0.6272&0.8048&0.7158&0.7911&0.7998&86,754\\BART-MIA&0.9064&0.6496&0.7782&0.9054&0.7375&382,606\\
    \hline
    \end{tabular}}
    \begin{tablenotes}
\footnotesize
\singlespacing
\item Note: We report standard measures of prediction accuracies (by column) for different methods we train (by row). For details on how prediction accuracies are constructed, see Appendix C. Any observation is a firm-year present in the sample. All methods but BART-MIA do not train or test on observations when at least one predictor is missing. Hence, a larger number of observations in testing BART-MIA.
\end{tablenotes}
\end{table}

From Table \ref{tab:horse_race}, we immediately notice that BART-MIA outperforms other methods with an ROC equal to $0.9054$, a value that is considerably higher than in the case of other methods. In fact, BART-MIA is in general more able than others to predict both exporters and non-exporters, with a Balanced Accuracy of $0.77$. 

Yet, when we look at Specificity \textit{vis \'a vis} Sensitivity values, we realize it predicts relatively better non-exporters rather than exporters. The reason is that the boost in overall prediction accuracy by BART-MIA is largely due to an efficient use of the non-random missing values on smaller firms reporting incomplete financial accounts. See also the specific robustness checks performed in \ref{sec: robustness}. As largely expected, smaller firms with partial information are also the ones that are more likely to be classified as non-exporters, because: i) larger size is more likely associated with an export status, and ii) smaller firms do not have to report financial information as complete as it is required to bigger companies. 

Since BART-MIA  is able to include the \textit{missingness} of any single feature as an additional predictor (i.e., as yet another \textit{branch} of the regression tree), we understand why it outperforms other methods, which instead simply drop from computation companies that have any missing values in predictors. 

Finally, a simple comparison between the accuracy of BART and the one of BART-MIA allows us to quantify what is the gain in considering the predictive power of missing values. Overall, we observe a $14.4\%$ increase in ROC, which we take as our baseline measure of prediction accuracy. We will further discuss the trade-off between Specificity and Sensitivity once we challenge our results in Section \ref{sec: discontinuous}. Suffice it to say here that, in general, predicting true exporters is made difficult by the presence of temporary trade, i.e., when firms export in some years and not in others, thus breaking the time series.

\subsection{Predictions}

\begin{figure}
\caption{Distributions of exporting scores of non-exporters after BART-MIA}
\label{fig:exporting_score}
\centering
{\includegraphics[width=0.7\textwidth]{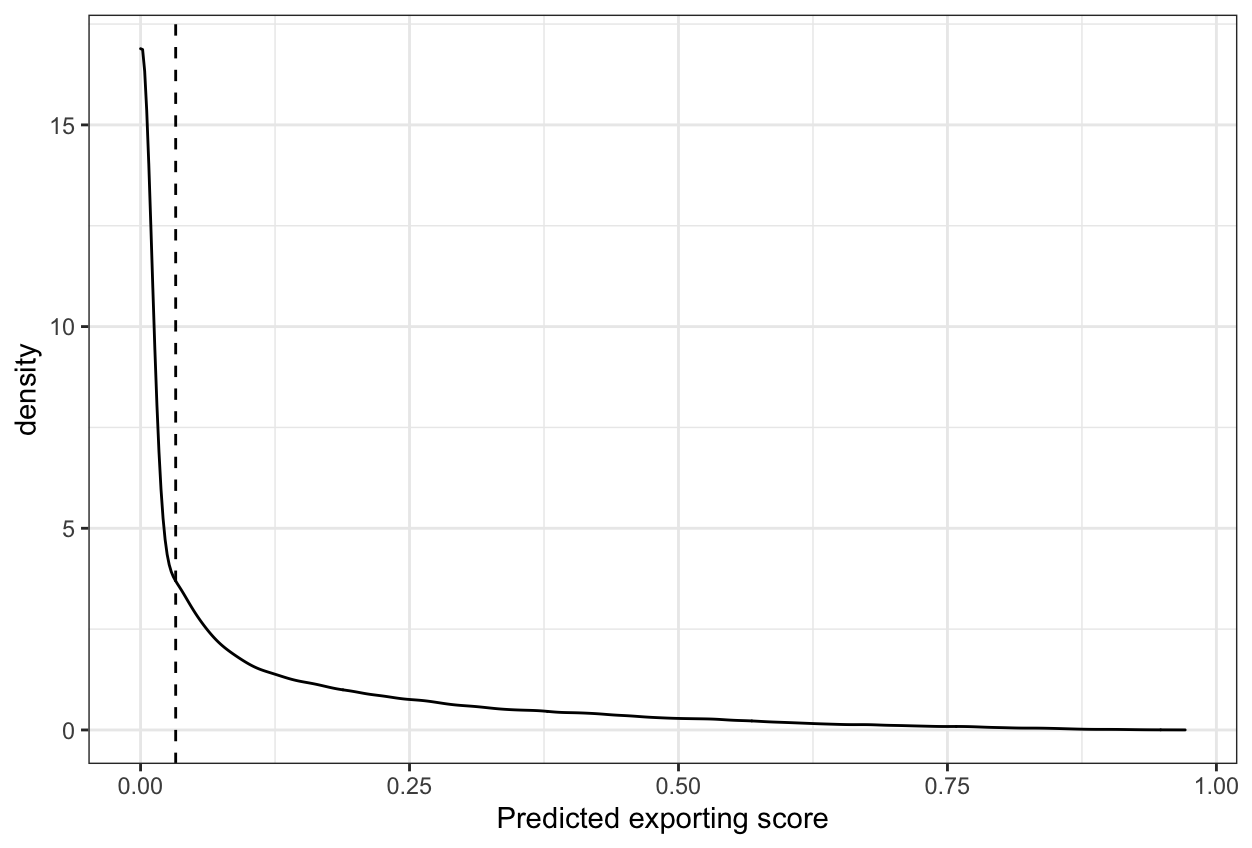}}
 \begin{tablenotes}
\footnotesize
\singlespacing
\item Note: We report the distribution of the score after implementing BART-MIA on the entire sample and selecting all non-exporting firms. The vertical line identifies the median non-exporting firm.
\end{tablenotes}
\end{figure}

In Figure \ref{fig:exporting_score}, we report the entire distribution of predicted scores for non-exporters that we obtain from our baseline BART-MIA. Without any selection threshold, these are the values that one could consider for evaluating how far a company is from export status. What is relevant to observe here is that the distribution is much skewed, hence the majority of non-exporters in France is located on a thick left tail, thus far from being able to propose on foreign markets. Briefly, the distribution of scores that we obtain here is consistent with the idea of firm heterogeneity that we take from trade literature, as introduced in Section \ref{sec: literature}. In other words, only a relatively small number of non-exporters is proximate to reaching the right tail's goal. The observation that firms are heterogeneous also in exporting scores is relevant for taking informed policy decisions that we discuss in Section \ref{sec: scoring}.

\newpage
\section{Robustness checks}\label{sec: robustness}

So far, we adopted a relatively standard $80-20$ random partition of the firms in the sample at our disposal when training our model \citep{athey2021using}. Therefore, our first concern here is to cross-validate our choice by repeating the prediction exercise other four times with a similar random partition. We want to check that our high prediction accuracy is not due to a fortunate selection of the training-and-testing partition. Any time, we train on a random $80\%$ of the dataset that we consider as in-sample information, then we test the accuracy of our predictions on the rest $20\%$, which we take as out-of-sample information. We show in Table \ref{tab:bart_mia_cv} how we obtain similar performance scores across all exercises, and we pick BART-MIA once again as the most predictive algorithm. We conclude that previous results had not been driven by a specific selection of training \textit{vis \'a vis} testing data.

Our second concern is that prediction accuracies are robust to different definitions of exporters. So far, we defined an exporter as any firm with positive exporting revenues. Here, we will define an exporter as a firm whose export share over total revenues is higher than a specific minimum threshold, to make our results robust to the presence of so-called \textit{passive exporters}  \citep{geishecker2019one}, i.e., domestic firms that engage in one-off exporting events.

Appendix Table \ref{tab: bart_mia_percentiles} shows prediction accuracies after we run simulations by excluding from the category of exporters those firms that report export shares lower than the first, second, and fifth percentile, respectively. Prediction accuracies are similar in magnitude to those of our benchmark definition. Latter evidence suggests that baseline predictions are not affected by the presence of a few less proactive firms.

A third concern we have is to verify the robustness to changes in predictors. Our problem here is whether we could obtain similar prediction accuracy with a minor effort, once neglecting variables that contribute with a relatively little predictive power. For this purpose, we perform a Logit-LASSO exercise before running again the models described in \ref{sec: methods}. As in standard applications \citep{ belloni2017program}, the Logit-LASSO selects a subset of best predictors (in our case, $23$ out of $52$) to contribute relatively more to predict export status.  Once again, BART-MIA outperforms other statistical learning techniques. However, when we perform BART-MIA including only such a subset of predictors, we obtain lower accuracy than baseline results, as reported in Appendix Table \ref{tab: subset_predictors}. Yet, we gather there is no reason to exclude available predictors despite the high cross-correlations we observed in Figure \ref{fig: corr_mat_pred}. 

A fourth concern we have is to check whether the time of training and testing matters for predictions. So far, we considered firms and their export status throughout the entire period at our disposal, between 2010 and 2018. In  Appendix Table \ref{tab:bart_mia_annual}, we train and test our predictive model separating each year. It is evident how predictions do not change dramatically over the timeline.

A fifth concern is that performance measures are robust to different probability thresholds for predicting the exporting status. In baseline analyses, we adopt a quite standard cut-off value set at $0.5$ to separate exporters and non-exporters in prediction. Yet, we know that exporting is a relatively rarer event than non-exporting, and our prediction accuracies can suffer from a bias. The choice of the threshold is, indeed, crucial for the computation of most prediction accuracies because the values in Table \ref{tab:horse_race} are threshold-specific. For a similar case in trade literature, see \cite{Baieretal2014}. Here we want to check that a different threshold does not alter the ranking of methodologies obtained by comparing prediction accuracies in Table \ref{tab:horse_race}. Therefore, in appendix Table \ref{tab: optimal_cutoff}, we show how performance measures vary when we choose, for each model, the optimal cut-off value obtained following \cite{liu2012classification}, who aims at maximizing the product of sensitivity and specificity. When an optimal threshold is set, the evidence of BART-MIA superiority is even more striking as it outperforms the others by all measures of prediction accuracy except for PR. We will discuss in section \ref{sec: discontinuous} how the latter is negatively affected by the presence of discontinuous exporters. Note, however, that both PR and ROC are not affected by the change in cut-off values because they are independent of thresholds by construction. The latter is also the reason why we consider them as baseline measures of performance.

A final concern is that baseline predictions improve mechanically only because the sample size is bigger in BART-MIA than in other exercises. In fact. we want to investigate whether improvements actually come from missing values. For our purpose, we perform two different exercises: i) we add \textit{ex ante} a predictor to our original set that catches the relative \textit{missingness} of financial information at the firm-level; ii) we impute missing values on single predictors based on median values available as from other companies' financial accounts. From a combined reading of both exercises, we better understand the role of \textit{missingness}.


Results for the latter exercise are reported in Appendix Table \ref{tab: imputed_data}. Interestingly, prediction accuracies do increase overall for all methods after predictors' imputation, although classification trees (BART\footnote{At this stage, computing BART-MIA or BART is equivalent, since we filled in missing value with imputations. The BART-MIA won't find any missingness, and won't include missing values among predictors, thus reversing to a more traditional BART procedure}, Random Forest), perform relatively better along the different segments of the distribution (ROCs are 0.907 and 0.905, respectively).  Eventually, when we check for the relative importance of a predictor on \textit{missingness}, we find that it is always selected as the best predictor no matter what procedure we choose. We conclude that missing values do have a prediction power, yet our baseline BART-MIA better catches their role without introducing unnecessary data manipulation.

    

\*

Eventually, we consider useful also reporting Spearman's rank correlations in Table \ref{tab:spearman2}, to test whether rankings in predictions are sensitive to the choice of predictive models in Table \ref{tab:horse_race}. Please note how, by construction, the Spearman's rank correlations can be performed only on the subset of the data where every technique obtains predictions. 

As a matter of fact, we get relatively high rank-correlations across predictive models with a minimum of $0.87$ and a maximum of $0.96$. In general, models do not dramatically alter the relative positions of firms on the distribution of predictions. Interestingly, please note that rank-correlation between the simpler BART and the BART-MIA is about $0.92$. Although the latter is just a variant of the first with \textit{missingness} of values as an additional feature, the rankings in predictions are different. The latter is a significant result that allows us to further qualify the difference between the simpler BART and its variant. The bottom line is that information from firms with missing values in predictors allows BART-MIA to identify different thresholds on predictors' distributions, which in turn change the relative positions of firms on the distribution of predictions.

\begin{table}[htpb]
\centering
\caption{Spearman's rank correlations of predicted probabilities from different models}
\label{tab:spearman2}
\resizebox{0.85\textwidth}{!}{%

\begin{tabular}{l|ccccc}
\hline
\hline
& LOGIT & LOGIT-LASSO & Random Forest & BART & BART-MIA\\
\hline
LOGIT & 1     &  0.9657 &   0.8773&0.8841&0.9012 \\
LOGIT-LASSO &  & 1 & 0.8925&0.9030&0.9118 \\
Random Forest &  &  & 1 & 0.9112 & 0.9167\\
BART & & & & 1 &0.9179\\
BART-MIA & & & & & 1\\
          \hline
\end{tabular}}
\begin{tablenotes}
\item Note: We report a Spearman's rank correlation among out-of-sample predictions to show how rankings in export status are sensitive to changes in predictive models. All models, including BART-MIA, are thus trained and tested on the same observations.
\end{tablenotes}

\end{table}

\section{Sensitivity to temporary trade} \label{sec: discontinuous}
We investigate in this section the sensitivity of our results to the presence of discontinuous exporting activity, i.e., when firms engage in trade relationship that are temporary \citep{BekesMurakozy}. Indeed, the biggest challenge we face when predicting exporters is that firms can export in some years and then lay idle for a while before re-proposing (or not) on foreign markets. This is especially true for smaller firms or for firms that are specialized in manufacturing capital goods. Thus, our prior is that discontinuity is not at random; it could be correlated with some firms' attributes, and our previous predictions could be therefore sensitive to the relevance of temporary trade within our sample. 

For our purpose, we perform separate checks by classifying firms into five categories:
\begin{enumerate}
    \item firms that always export, which we call \textit{constant exporters};
    \item firms that never export, which we call \textit{non-exporters};
    \item firms that start exporting at some period $t$ and always export afterwards, which we call \textit{switching exporters};
    \item firms that export in all periods until $t$ and never export afterwards, which we call \textit{switching non-exporters}\footnote{Please note how we may have had more switching non-exporters if we were able to zoom out on a longer timeline. We cannot exclude that firms that do not export in our sample did so in previous unobserved periods. The latter is an element of imperfection that we cannot expunge from our prediction exercise.};
    \item \textit{discontinuous exporters}, which export with an irregular pattern with more than one gap along the timeline.
\end{enumerate}

Prediction accuracies are eventually reported in Table \ref{tab: bart_mia_firm_cat}, after testing out-of-sample our baseline BART-MIA algorithm. As expected, we observe that our predictive model performs quite well in separating constant exporters from non-exporters, since Sensitivity and Specificity are about $0.86$ and $0.95$, respectively\footnote{Please note that we cannot estimate other measures of prediction accuracy when we focus exclusively on either positive or negative outcomes. See Appendix C for a definition of different measures of prediction accuracies.}. On the other hand, predictions become relatively less accurate when we look at out-of-sample information on firms that show gaps along the timeline. In general, we still have acceptable accuracies as ROCs reach up to $0.86$ and $0.81$, respectively, in the case of \textit{switching exporters} and \textit{switching non exporters}. In line with our priors, the quality of predictions is proportional to the number of years that the firms actually exported. Predictions are more accurate when firms started (stopped) exporting sooner (later) in our data.\\

Finally, we focus on the category what we define discontinuous exporters, when firms have more than one break in the time series, entering and exiting the export status. In this case, at the bottom of Table \ref{tab: bart_mia_firm_cat}, we find that prediction accuracy reached a relatively lower albeit acceptable threshold (ROC: $0.80$). The accuracy is lower than the one obtained in predicting constant exporters and non-exporters. Interestingly, we do register that our procedure is less and less able to predict the export status in the case of firms that have less experience of foreign markets. This is however consistent with the idea that firms engaging in temporary trade may continue to do so systematically, hence their lower predictability on a year-by-year basis.

Eventually, a final sensitivity check to temporary trade is performed by introducing a more liberal definition of exporters proposed by \citet{BekesMurakozy}, according to whom only firms with at least four years of consecutive exporting can be actually considered as \textit{permanent exporters} vis \'a vis \textit{temporary exporters}. As largely expected, we find in Appendix Table \ref{tab: bart_mia_BM} that prediction accuracies for \textit{permanent exporters} are relatively higher (AUC: $0.849$; PR: $0.934$) than in the case of temporary exporters. In particular, the model fails at predicting the export status of temporary exporters, i.e., it reports a relatively lower true positives' rate, as shown by the low scores on sensitivity, PR and ROC. 

From our viewpoint, it makes sense that exporters with irregular exporting patterns represent intermediate cases somewhere between firms that always export and firms that never export. Therefore, classification algorithms struggle to separate intermediate cases on a binary outcome. Based on financial accounts, such firms can be seen neither as fit for exporting as constant exporters, nor as unfit as non-exporters. Yet, it is more likely that such intermediate cases are of less interest in policy applications because trade promoters or financial institutions need instead to understand whether a firm that never exported at all needs some support or not.

\begin{table}[H]
\caption{Prediction accuracies and temporary trade}
    \label{tab: bart_mia_firm_cat}
    \centering
    \resizebox{0.75\textwidth}{!}{%
    \begin{tabular}{lcccccc}
    \hline \hline
    Firm category &Sensitivity&Specificity&  Balanced&ROC &PR& Num.\\
     &&& Accuracy&&& Obs.\\[0.5ex]
    \hline

Constant Exporters&0.856&-&-&-&-&21,834\\
Non-exporters&-&0.951&-&-&-&158,625\\
&&&&&\\[-2ex]
\hline
Switching to export&0.629&0.849&0.739&0.864&0.764&15,084\\
\textit{Since $t_0$}&\textit{0.749}&\textit{0.682}&\textit{0.716}&\textit{0.794}&\textit{0.954}&\textit{1,980}\\
\textit{Since $t_1$}&\textit{0.729}&\textit{0.694}&\textit{0.712}&\textit{0.808}&\textit{0.914}&\textit{1,296}\\
\textit{Since $t_2$}&\textit{0.711}&\textit{0.751}&\textit{0.731}&\textit{0.838}&\textit{0.888}&\textit{1,179}\\
\textit{Since $t_3$}&\textit{0.618}&\textit{0.806}&\textit{0.712}&\textit{0.832}&\textit{0.821}&\textit{1,215}\\
\textit{Since $t_4$}&\textit{0.582}&\textit{0.796}&\textit{0.689}&\textit{0.812}&\textit{0.73}&\textit{1,323}\\
\textit{Since $t_5$}&\textit{0.585}&\textit{0.819}&\textit{0.702}&\textit{0.823}&\textit{0.638}&\textit{1,683}\\
\textit{Since $t_6$}&\textit{0.463}&\textit{0.835}&\textit{0.649}&\textit{0.804}&\textit{0.45}&\textit{2,187}\\
\textit{Since $t_7$}&\textit{0.262}&\textit{0.903}&\textit{0.583}&\textit{0.792}&\textit{0.251}&\textit{4,221}\\

\hline

&&&&&\\[-2ex]
Switching to non-export&0.599&0.802&0.7&0.819&0.786&27,891\\
\textit{Until $t_0$}&\textit{0.269}&\textit{0.81}&\textit{0.539}&\textit{0.643}&\textit{0.152}&\textit{3,915}\\
\textit{Until $t_1$}&\textit{0.376}&\textit{0.745}&\textit{0.561}&\textit{0.65}&\textit{0.291}&\textit{2,511}\\
\textit{Until $t_2$}&\textit{0.419}&\textit{0.725}&\textit{0.572}&\textit{0.689}&\textit{0.443}&\textit{2,124}\\
\textit{Until $t_3$}&\textit{0.479}&\textit{0.737}&\textit{0.608}&\textit{0.733}&\textit{0.599}&\textit{2,412}\\
\textit{Until $t_4$}&\textit{0.508}&\textit{0.815}&\textit{0.662}&\textit{0.816}&\textit{0.757}&\textit{2,844}\\
\textit{Until $t_5$}&\textit{0.563}&\textit{0.925}&\textit{0.744}&\textit{0.929}&\textit{0.924}&\textit{5,409}\\
\textit{Until $t_6$}&\textit{0.664}&\textit{0.843}&\textit{0.754}&\textit{0.877}&\textit{0.931}&\textit{3,996}\\
\textit{Until $t_7$}&\textit{0.742}&\textit{0.813}&\textit{0.778}&\textit{0.874}&\textit{0.97}&\textit{4,680}\\
&&&&&\\[-2ex]
\hline

Discontinuous exporters&0.547&0.807&0.677&0.796&0.686&85,023\\
\textit{export experience: 1 year}&\textit{0.216}&\textit{0.873}&\textit{0.544}&\textit{0.686}&\textit{0.171}&\textit{19,152}\\
\textit{export experience: 2 years}&\textit{0.313}&\textit{0.823}&\textit{0.568}&\textit{0.702}&\textit{0.334}&\textit{12,816}\\
\textit{export experience: 3 years}&\textit{0.387}&\textit{0.796}&\textit{0.592}&\textit{0.718}&\textit{0.483}&\textit{10,962}\\
\textit{export experience: 4 years}&\textit{0.478}&\textit{0.736}&\textit{0.607}&\textit{0.719}&\textit{0.595}&\textit{8,910}\\
\textit{export experience: 5 years}&\textit{0.519}&\textit{0.74}&\textit{0.63}&\textit{0.753}&\textit{0.72}&\textit{9,297}\\
\textit{export experience: 6 years}&\textit{0.593}&\textit{0.721}&\textit{0.657}&\textit{0.755}&\textit{0.808}&\textit{8,460}\\
\textit{export experience: 7 years}&\textit{0.662}&\textit{0.7}&\textit{0.681}&\textit{0.774}&\textit{0.886}&\textit{7,758}\\
\textit{export experience: 8 years}&\textit{0.757}&\textit{0.658}&\textit{0.708}&\textit{0.781}&\textit{0.951}&\textit{7,668}\\
&&&&&\\[-2ex]
\hline
All sample & 0.6491 & 0.9080 & 0.7785&0.9048&0.7383&308,457
\end{tabular}}
\begin{tablenotes}
\footnotesize
\singlespacing

\item Note: We report prediction accuracies after BART-MIA for firms with different exporting patterns. For switching-exporters and switching-non-exporters we identify the year when they are observed changing status, i.e., the year when the firm passes from never exporting to always exporting, and vice versa. For discontinuous exporters we distinguish by number of exporting years over the sample timeline.
\end{tablenotes}

\end{table}

\section{Interpretability of predictors}\label{sec: coefficients}

\subsection{Influence on predictions}

In line with our empirical strategy, we focused so far on prediction accuracy while neglecting the role of single predictors. We discussed in Section \ref{sec: strategy} how our choice is driven by the necessity to maximize prediction accuracy; therefore we have been using an as complete as possible list of predictors, even though we are aware that we carried on with us a compound of endogenous variables that are highly cross-correlated, as commented after Figure \ref{fig: corr_mat_pred}. 

What we want to do now is to show how predictors do have different influence on the outcome, and we can still discuss their influence on predictions without implicating any causality. On the contrary, the internal validity of our 'influential predictors' is to us more important than an external validity. They are relevant because we can interpret them in relationship with the specific prediction exercise we want to comment. If we consider a different sample, those 'influential predictors' will be almost certainly different.

Our baseline method for the interpretability of a BART-MIA exercise is called Variable Inclusion Proportions (VIP)\footnote{For a different choice of methods to catch the relative importance of predictors, see also \citet{Joseph2020} and the case of neural networks.}. The Variable Inclusion Proportion for any given predictor represents the proportion of times
that variable is chosen as a splitting rule out of all splitting rules among the posterior draws
of the sum-of-trees model \citep{kapelner2013bartmachine}.  It is computed as follows: (1) Across all $q$ trees in the ensemble (\ref{eq: tree}), we examine the set of predictor variables used for each splitting rule in each tree; (2) For each sum-of-tree model we compute the proportion of times that a split using $x_i$ as a splitting variable
appears among all splitting variables $\bX$ in the model; (3) with $K$ being the number of the sum-of-tree models $f^*_k$, drawn from the posterior distribution $\mathbb{P}(\mathcal{T}_1^\mathcal{M},...,\mathcal{T}_m^\mathcal{M},1 | \Phi(Y))$, and $z_{ik}$ being the proportion of all splitting rules that use the $i^{th}$ component of $\bX$ in model $f^*_k$, the Variable Inclusion Proportion is computed as
\begin{equation}
    v_i=\frac{1}{K} \sum_{k=1}^K z_{ik}
\end{equation}

\begin{figure}
    \centering
       \caption{Variable inclusion proportions after BART-MIA}
       \resizebox{0.9\textwidth}{!}{%
    \includegraphics[width=\textwidth]{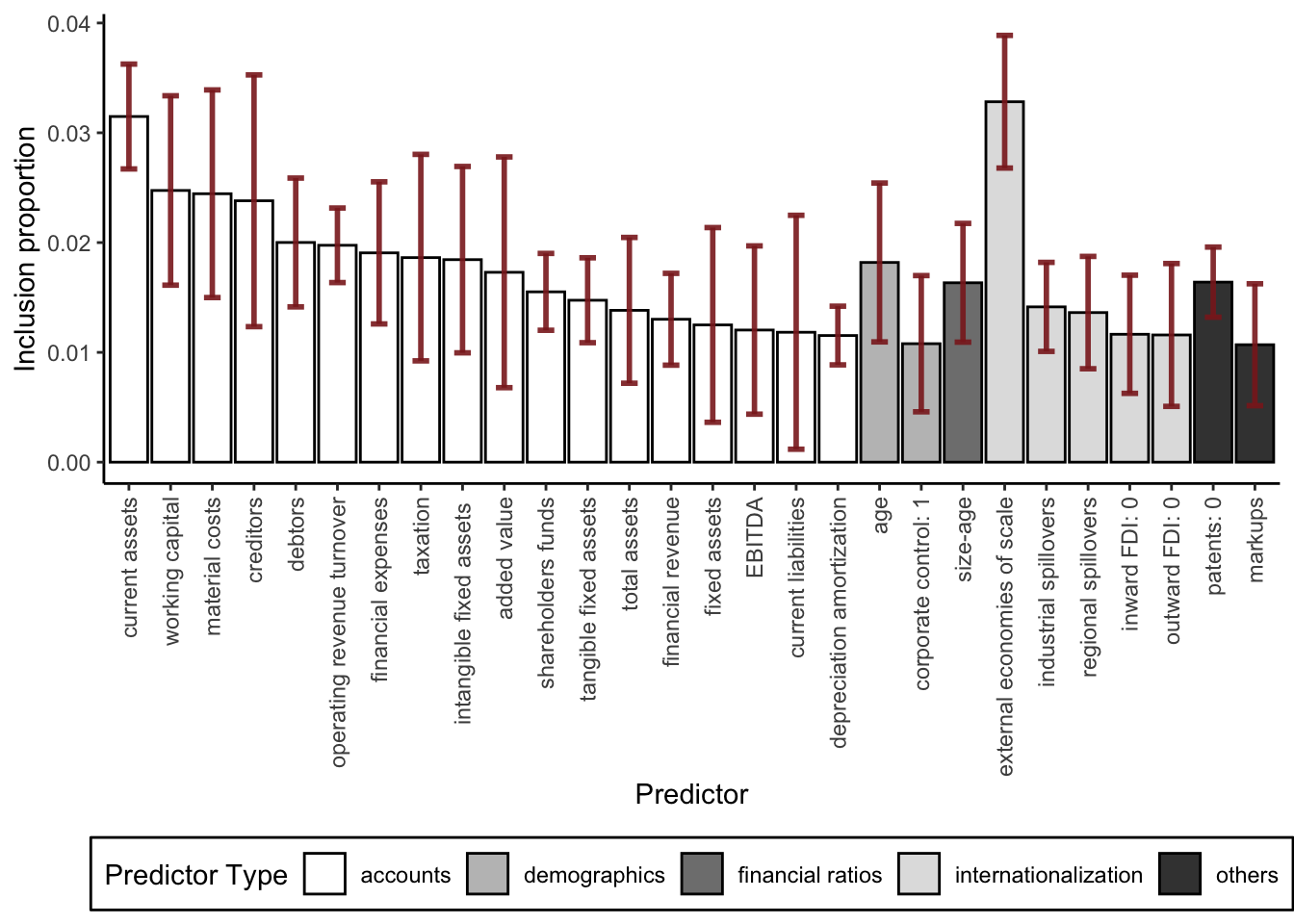}
    \label{fig:incl_prop}}
            \begin{tablenotes}
\footnotesize
\item Note: We report Variable Inclusion Proportions (VIPs), i.e., the proportion of times each predictor is chosen for a splitting rule in BART-MIA. Of all the predictors in baseline, we visualize only those with a VIP higher than $1\%$. Red bars represent standard deviations of inclusion proportions obtained by replicating five different times the BART-MIA on the same random training set.
\end{tablenotes}
\end{figure}

Thus, we report in Figure \ref{fig:incl_prop} a visualization of the VIPs accompanied by a standard deviation that is computed after running five different random tests. Please note how averaging across multiple trials allows us to improve the stability of estimates, as suggested by \cite{kapelner2013bartmachine}. For the sake of visualization, we report in Figure \ref{fig:incl_prop} only those predictors that register a VIP equal or higher than $1\%$.

When we look at Figure \ref{fig:incl_prop}, we document that the best predictor in our baseline exercise is the proxy we use for the existence of external economies of scale, which indicates the presence of other firms in the same industry and in the same region, as suggested by \cite{bernard1995exporters}. Once again, we want to stress that since we are in a pure prediction framework, we cannot say whether external economies of scale, measured in this way, are an actual determinant of export status. We cannot exclude reversal causality. On the one hand, it is indeed possible that local spillovers help neighbouring firms to start exporting after, for example, sharing infrastructures or intangible knowledge about foreign markets. On the other hand, it is possible that firms in industries at a comparative advantage locate in geographical proximity before becoming exporters. In any case, it is beyond the scope of our analysis to unravel the endogeneity of this specific relationship or any other we know we have among predictors and the outcome. Suffice it to say that the industrial concentration of exporting firms in a region of France is a good albeit not unique predictor of export status for the representative firm located in that area.

Notably, we observe in Figure \ref{fig:incl_prop} how original accounts altogether provide an important contribution to predict export status. Yet, no single predictor contributes more than $4\%$ in any of the tests we performed. Besides financial accounts, business demography has predictive power: firm age has an inclusion proportion higher than $2\%$. It also makes perfect sense that the activities of multinational enterprises play a role in export status. Being either a foreign subsidiary (inward FDI) or owning a subsidiary abroad (outward FDI) affects the probability of exporting. As expected, the ability to innovate and register patents is also related to the likelihood of becoming an exporter. 

Eventually, we want to bring the attention on the absence of Total Factor Productivity (TFP) in Figure \ref{fig:incl_prop}, which we however included following the methodology by \cite{ackerberg2015identification}. Although TFP is a much-studied determinant of export status, we do not find it to be among the most relevant predictors in a machine learning exercise. Our educated guess is that the role of TFP is already captured by the sample variation in raw financial accounts that are also needed to compute it as a residual from a firm-level production functions (turnover, costs of materials, employees, etc.).

\subsection{Internal vs. external validity}\label{sec: internal validity}

We cannot presume external validity when we want to interpret the VIPs that constitute a posterior probability that the  variable $x_k$ has a \textit{real effect}, defined as the
impact of some linear or nonlinear association, on the response variable \citep{bleich2014variable}. Variables selected through VIPs would be almost certainly different if we consider a different country or region than France. Yet, we argue that the relevance of VIPs resides in their internal validity, given the peculiarity of each predictive exercise. For example, one could compare across different regions how the relative importance of predictors changes and use that information to take more solid policy decisions. 

\begin{figure}[htp]
 \caption{Variable inclusion proportions in \^{I}le-de-France \textit{versus} the rest of France}
 \label{fig: ext_validity}
\subfloat[VIPs \^{I}le-de-France]{%
  \includegraphics[clip,width=0.9\columnwidth]{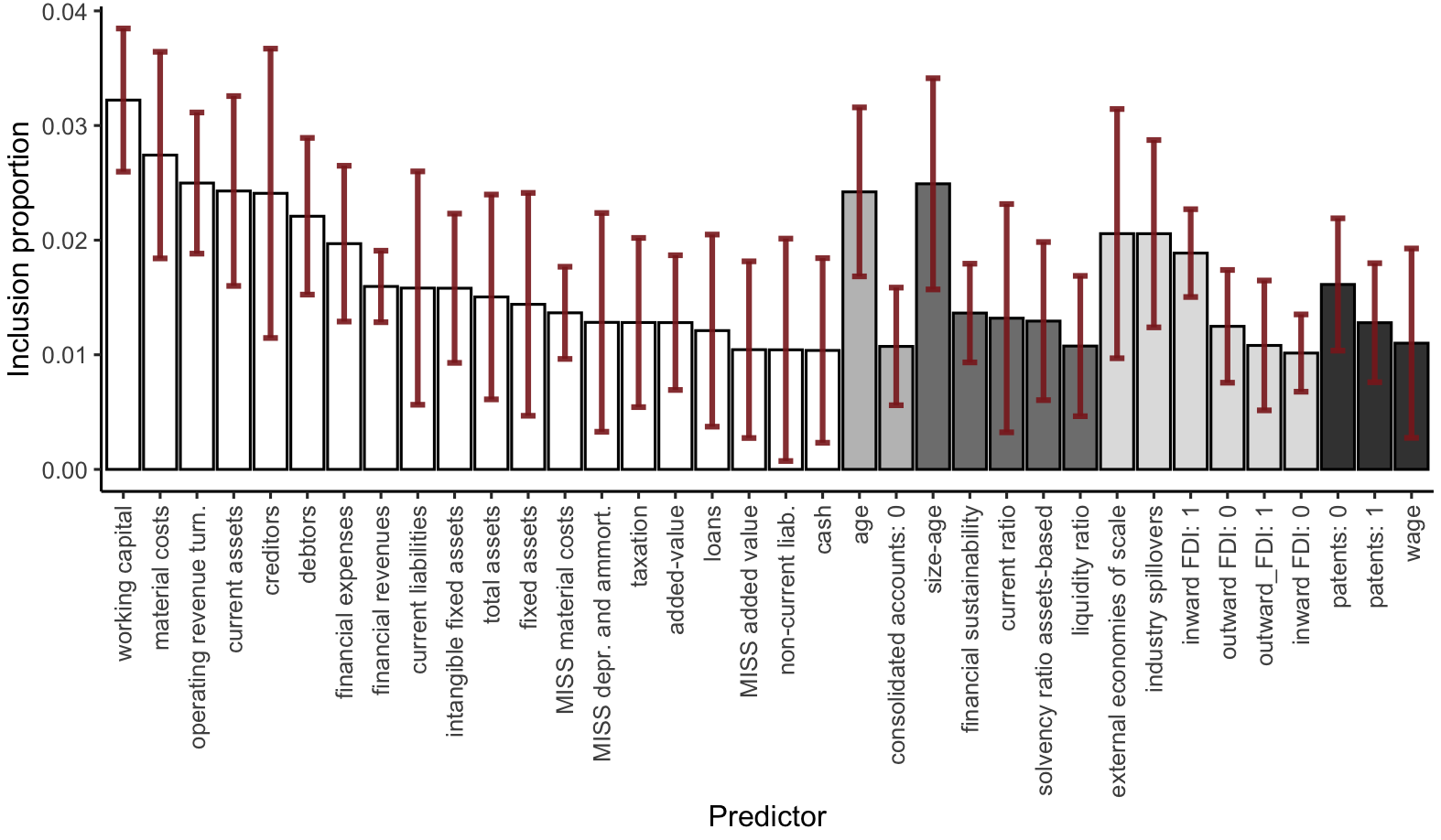}%
}

\subfloat[VIPs France, excluding \^{I}le-de-France]{%
  \includegraphics[clip,width=0.9\columnwidth]{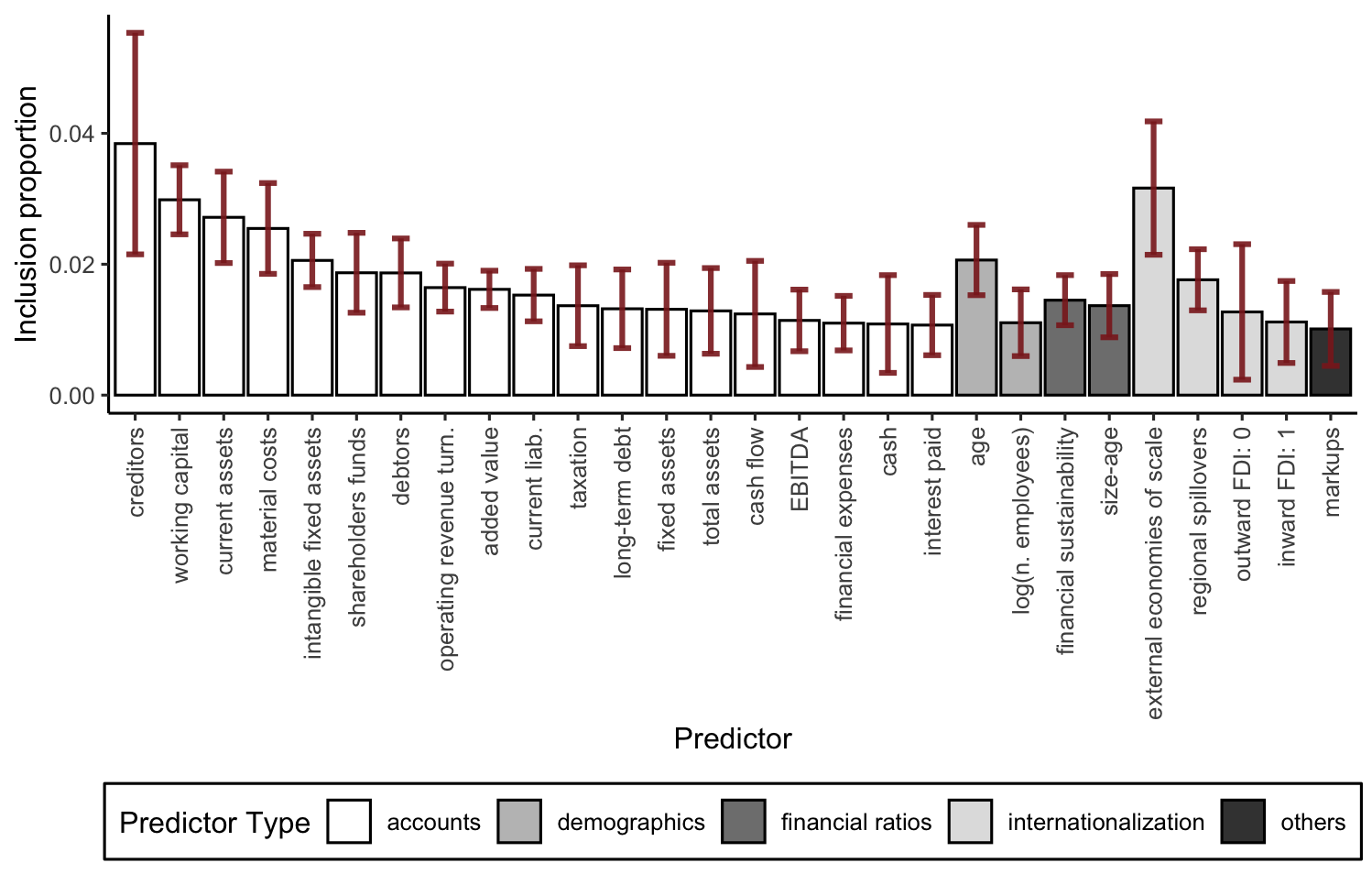}%
}

 \begin{tablenotes}
\footnotesize
\item Note: We report Variable Inclusion Proportions (VIPs) in (a)\^{I}le-de-France, (b) in all France \textit{excluding}\^{I}le-de-France. Of all the predictors in baseline, we visualize only those with a VIP higher than $1\%$. Red bars represent standard deviations obtained by replicating five different times the BART-MIA on the same random training set.
\end{tablenotes}

\end{figure}

To make our point, we replicate our exercise after separating \^{I}le-de-France from the rest of the country. We show only VIPs for both subsets in Figure \ref{fig: ext_validity}. 

We observe that not only the set of influential predictors differs, but also that the relative importance of predictors changes from one exercise to the other. This hints at the presence of locally different dynamics. For example, note how the predictor \textit{(log of) number of employees} is selected in the sample excluding \^{I}le-de-France, but not in \^{I}le-de-France, where there is possibly more homogeneity in terms of firm size. In contrast, the predictor \textit{patent} is influential in \^{I}le-de-France, but not elsewhere, possibly indicating that in the first there is a comparative advantage in more innovative activities that have the potential to reach foreign markets. \textit{Prima facie}, the latter evidence is consistent with the prior knowledge we have about the landscape of the French economy.

\section{How to use exporting scores}\label{sec: scoring}

In this Section, we provide examples of possible applications of exporting scores as either indicators for trade credit or a useful tool for assessing the trade potential of regions and industries. Based on the prior knowledge that exporters and non-exporters are statistically different across financial attributes, we use in-sample information to predict out-of-sample capability to export. Thus, it is immediate to build a continuous indicator that provides an exporting score based on previous baseline predictions, to indicate the potential of companies to successfully propose on foreign markets, i.e., their distance from export status. We visualize our intuition in Figure \ref{fig: intuition}, where we reported predictions for non-exporters obtained after the baseline BART-MIA in Figure \ref{fig:exporting_score}.

Briefly, we can get a basic and simple export (probabilistic) score for any out-of-sample non-exporting $i$th firm, in the form:

\begin{equation}\label{eq: distance}
    {distance}_{i}= 1 - {Pr}(Y_{i}=1 \:|\: \bX_{i}=x)
\end{equation}

which is by definition bounded in a range $(0,1)$, and made conditional on the set of predictors, $\bX_{i}$, as from previous exercises.

\subsection{Exporting scores and trade credit}

Exporting requires routine access to financial resources, thus well-functioning ﬁnancial markets are crucial to support the daily activity of exporters \citep{Manova2012}. Since they incur high fixed costs to access distant foreign markets, exporters depend relatively more on external resources than domestic producers. Therefore, the presence of financial market imperfections constrains trade opportunities, all the more when firms are heterogeneous in the ability to provide collateral \citep{Chor&Manova2012}. 

Against this background, national and international agencies establish trade promotion programs to fill the gap in financial markets' imperfections and develop skills that help catch business opportunities on global markets \footnote{A variety of services are provided to firms that apply for trade support programs, ranging from training to financial resources.  International organizations specifically support firms in less advanced countries to fill the gap in global markets. See, for example, the experience of the Inter-American Development Bank and the International Trade Center. }. Eventually, export promotion programmes are effective tools in helping firms reach new destination countries and introduce new differentiated products \citep{Volpe&Carballo2010}. They facilitated the recovery after the global recession of 2009 \citep{VanBiesebroecketal2016}.
\\

In the case of credit risk, scholars and practitioners have developed several tools to reduce the informative gap between borrowers and lenders from disclosed financial accounts. The main idea has been to check how far a company is from a situation of financial distress using some combination of financial ratios \footnote{For example, Z-scores \citep{altman1968financial, altman2000predicting} and Distance-to-Default \citep{merton1974pricing} have been used as standard tools to assess a firm's viability on a combination of financial accounts. Recent advances in predictive models for bankruptcies also include machine learning methods. See, for example, \cite{zombie}.}. Yet, the case of would-be exporters is peculiar, as the problem is to monitor which companies are solid enough to get their products on foreign markets, and how much they need to bridge the gap.

\*

To illustrate our idea, we perform here back-of-the-envelope estimates to predict how much capital and cash resources may be needed by a company to become fit for exporting. We classify firms in different \textit{risk categories}, i.e., categories based on a partition of the distribution of exporting scores obtained in Figure \ref{fig:exporting_score}. For simplicity, let us consider all firms included in a decile of predictions as belonging to the same \textit{risk category}. Obviously, the higher the distance from export status, $1 - {Pr}(Y_{i})$, the higher the risk for trade credit. We obtain symmetric segments of length equal to $0.1$, i.e., about ten percentage points of lower risk in each category when approaching export status. Therefore, we can run the following simple specification:

\begin{equation}\label{eq: linear}
    \log Y_{it} =\beta_{0} + \sum_{risk=1}^{10} \theta_{risk} + \beta_{1} x_{it} +  + \phi_{t} + \delta_{s}+\eta_{r} + \epsilon
\end{equation}

where $Y_{it}$ is either cash resources or fixed assets for firm $i$ at time $t$, and $x_{it}$ is its firm-level size. We will always control for time ($\phi_{t}$), four-digit NACE sector ($\delta_{t}$), and two-digit NUTS region ($\eta_{r}$) fixed effects. We cluster standard errors at the firm level.
\*

\begin{figure}[H]
    \centering
    \caption{Premia on relevant firm dimensions across exporting scores}
    \label{fig: coefficients}
    \includegraphics[width=0.9\textwidth]{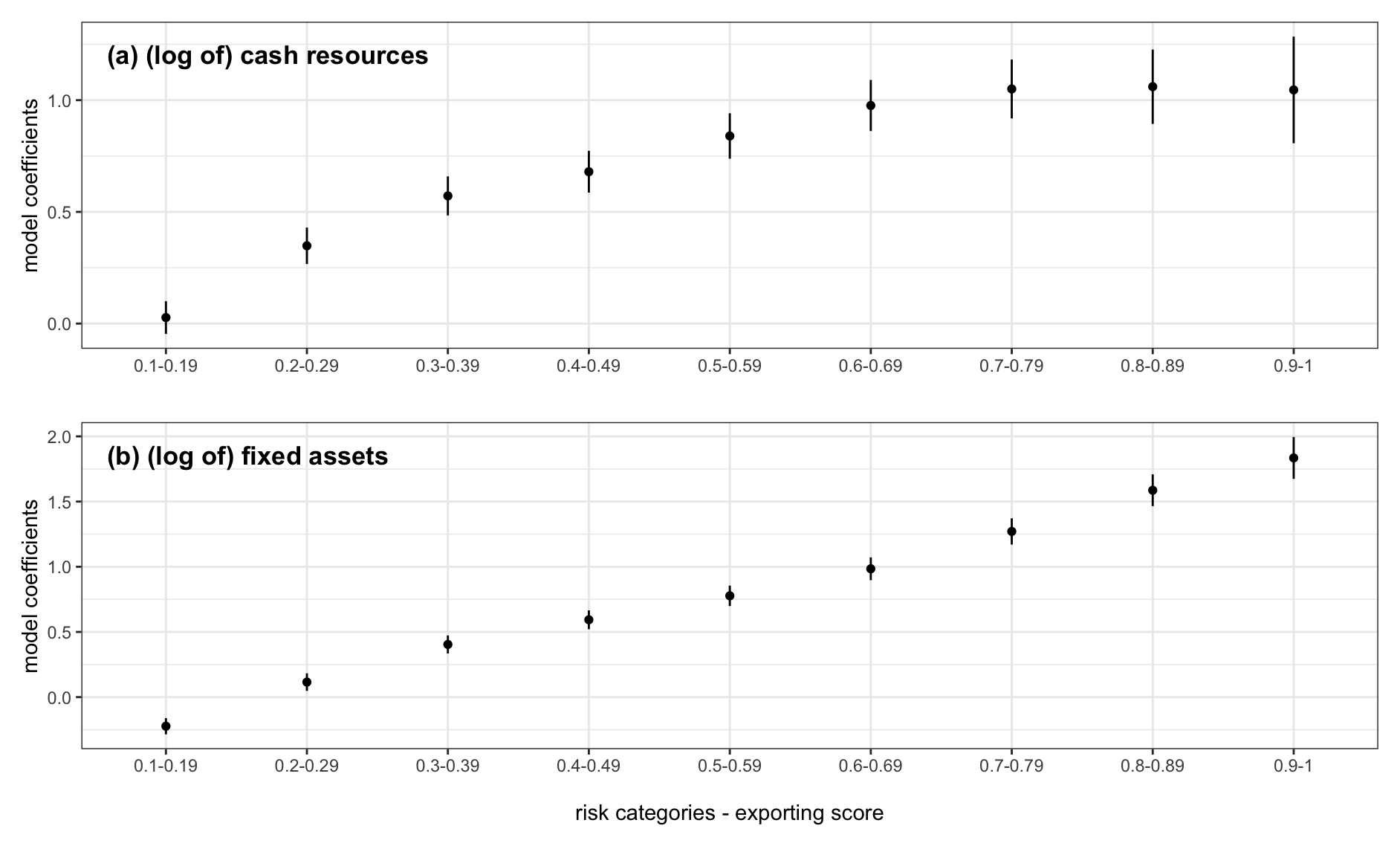}
     \begin{tablenotes}
\footnotesize
\singlespacing
\item Note: Fixed effects on segments of exporting scores after linear regressions where the outcomes are (log of) cash resources and (log of) fixed assets, respectively. We always control for firm size, NUTS 2-digit regions, NACE 2-digit industries, and time fixed effects. Errors are clustered at the firm level.
\end{tablenotes}
\end{figure}

Crucially, our coefficients of interest are the ones on $\theta_{risk}$, as these are the risk classes we built on exporting scores. We report them in decreasing order of risk in Figure \ref{fig: coefficients} together with $99\%$ confidence intervals. Once we omit the first segment $[0, 0.09]$, the estimated intercepts of eq. \ref{eq: linear} will indicate (logs of) cash resources and fixed assets needed by a representative firm that is more distant from export status. To obtain what is on average needed by a firm in a \textit{risk category}, we predict (log) premia with respect to the baseline omitted first segment.

For example, the representative firm with exporting scores lower than $0.1$ operates with $exp(\hat{\beta_{0}}) = exp(11.6338) \approx 112,850$ euro of cash resources and $exp(\hat{\beta_{0}}) = exp(13.4027) \approx 661,790$ euro of fixed assets. Firms in the fifth category, when exporting scores are in a range $[0.4, 0.5)$, will need $exp({\hat \beta_{0} + \hat{\theta_{5}}}) = (11.6338 + 0.6797) \approx 222,690$ euro of cash resources and $exp({\hat \beta_{0} + \hat{\theta_{5}}}) = exp(13.4027 + 0.5933) \approx 1,197,800$ euro of fixed assets. To put it differently, we can say that a firm that is in a medium-risk category needs about $97\%$ more cash resources and about $81\%$ more fixed assets if compared with a firm with the lowest exporting scores.  

On the other hand, if we look at firms in a comfort zone with exporting scores in a range $[0.9, 1]$, we see that they operate with $exp({\hat \beta_{0} + \hat{\theta_{10}}}) = exp(11.6338 + 1.0459) \approx 321,160$ euro of cash and $exp({\hat \beta_{0} + \hat{\theta_{10}}}) = exp(13.4027 + 1.8348) \approx 4,145,360$ euro of fixed assets. Please note that the higher the probability that a firm starts exporting, the higher the cash resources and the capital expenses it needs. In the latter case, if we compare with average exporting scores in the fifth risk class, we find that medium-risk firms need $44\%$ more cash resources and up to $246\%$ more capital expenses to look like firms that have been classified under the lowest risk category.

In terms of trade credit, we observe that there is an increasing need for financial resources to climb risk categories and reduce the distance from export status. Based on predictions made on the experience of both exporters and non-exporters, a financial institution could evaluate whether it's worth the effort of investing in internationalization and, in case, how much resources a firm needs to reach its target.

\subsection{Exporting scores and trade potential}

Finally, we spend a few words to show how exporting scores can help in assessing the potential for expanding the set of exporters in a region or an industry, i.e., the potential for a trade extensive margin. 

Openness to international trade is a determinant of economic growth. Thanks to differential comparative advantages and economies of scale, consumers can gain from trade. Both developed and developing economies have benefited from integration into the global economy through export growth and diversification. Thus, export performance has been long used as yet another proxy for measuring countries' competitiveness by a consolidated tradition in economic literature and by international organizations \citep{Leamer&Stern, Richardson1971a, Richardson1971b, Gaulieretal2013}. 


To make our point, we follow a dartboard approach as in \cite{EllisonGlaeser1997} and propose location quotients in Figure \ref{fig:quotients}. See Appendix D for further details on computations. Regions with location quotients greater than one are the ones where potential exporters are more concentrated than what one would expect given the underlying distribution of manufacturing activities. Eventually, we do find a geographic pattern in Figure \ref{fig:quotients}, since non-exporters with the highest potential are mainly present in North-Eastern regions, while Southern regions and overseas territories lag behind in trade potential.




\begin{figure}
\caption{The potential for extensive margin across France}
    \label{fig:quotients}
    \centering
    \includegraphics[width=0.9\textwidth]{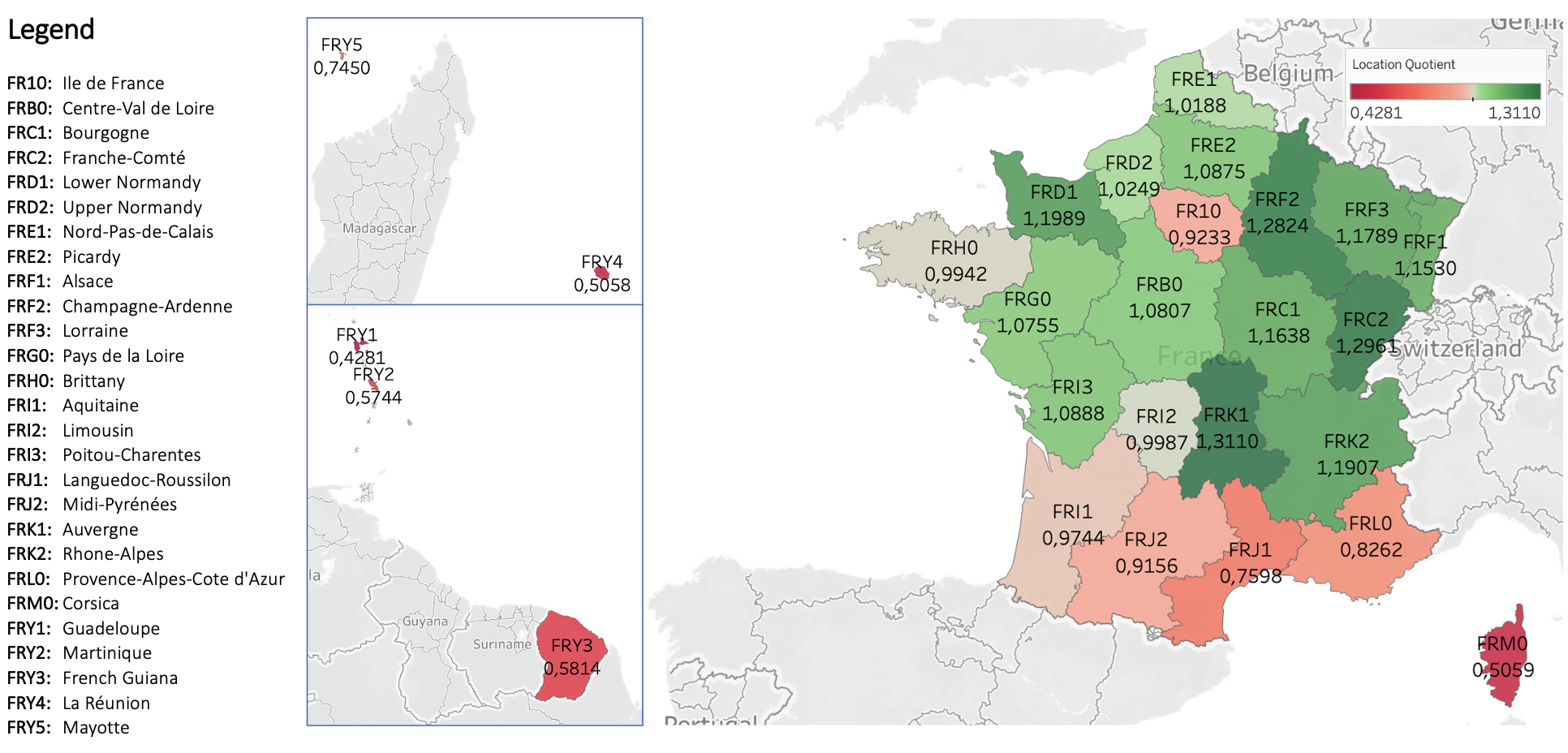}
        \begin{tablenotes}
\footnotesize
\singlespacing
\item Note: We report location quotients of non-exporters whose score is above the median in the national distribution. Regions with location quotients greater than one (lower than one) are those where potential exporters are more (less) concentrated than what one would expect given manufacturing density. Regions are reported in grey if location quotients are not statistically significant in a $90\%$ confidence interval. See Appendix D for details on the computation of location quotients.
\end{tablenotes}
\end{figure}


Eventually, more sophisticated analyses on the distribution of exporting scores in industries and regions can be performed to evaluate trade potential. For example, one could exploit the variation in time to understand how much competitive in trade a region or an industry is becoming. Also, one could compare across countries to check whether there is a different potential for trade beyond actual export performance. We believe any of them could be a useful tool in the kit of the analyst that aims at assessing the trade competitiveness of an economy.


\section{Conclusions}\label{sec: conclusions}

This paper exploits statistical learning techniques to predict the ability of firms to export. After showing how financial accounts convey non-trivial information to separate exporters from non-exporters, we propose predictions as a tool that can be useful for targeting trade promotion programs, trade credit, and assessing firms' trade potential. 

The central intuition is that exporters and non-exporters are statistically different in their financial structures since they have to sustain the sunk costs of gaining access to foreign markets, where regulations and consumer tastes differ. On this, we rely on established literature that relates firms' heterogeneity with self-selection into exporting activity. Thus, we train and test various algorithms on a dataset of French firm-level data from 2010-2018. Eventually, we find that the Bayesian Additive Regression Tree with Missingness In Attributes (BART-MIA) outperforms other models due to an efficient use of the non-random missing information on smaller firms reporting incomplete financial accounts. 

Notably, prediction accuracy is rather high, up to $90\%$, and robust to both changes in the definition of exporters and different training strategies. Interestingly enough, our framework allows handling cases of discontinuous exporters, as they show up as intermediate cases between permanent exporters and non-exporters. 


Eventually, we discuss how export predictions can be used as scores to catch the sustainability of firms' internationalization strategies and their creditability. For example, imitating what a financial institution would professionally do, we order firms along \textit{risk categories}. Thus, we show back-of-the-envelope estimates of how much cash resources and capital a firm would need to climb risk classes and become fit for exporting. In our case study, we show that a median French non-exporter needs up to $44\%$ more cash and $246\%$ more capital to reach full export status.

To conclude, we argue that exporting scores obtained as predictions from firm-level financial accounts can be yet another useful tool in the analyst kit to evaluate trade potential at different levels of aggregations. As we show in the case of France, for which we provide summary statistics where a high heterogeneity of trade potential is detected across regions.

\onehalfspacing
\setlength\bibsep{0.5pt}
\bibliographystyle{elsarticle-harv}
\bibliography{biblio.bib}

\newpage
{\centering
\section*{Appendix A: Data}

\setcounter{table}{0}
\renewcommand{\thetable}{A\arabic{table}}
\large{Table A1: Panel (A): List of predictors}\\
\begin{table}[htpb!]
\begin{tabular}{p{8cm}p{8cm}}
\hline
\textbf{Variable}& Description\\
\hline
Value Added, Depreciation, Creditors, Current Assets, Current liabilities, Non-current liabilities, Current ratio, Debtors, Operating Revenue Turnover, Material Costs, Costs of
Employees, Taxation, Financial Revenues, Financial
Expenses, Interest Paid, Number of
Employees, Cash Flow, EBITDA, Total Assets, Fixed Assets,
Intangible Fixed Assets, Tangible Fixed Assets, Shareholders' Funds, 
Long-Term Debt, Loans, Sales, Solvency Ratio, Working Capital
& Original financial accounts expressed in euro.\\
Corporate Control & A binary variable equal to one if a firm belongs to a
corporate group.\\
Dummy Patents & equal to 1 if the firm issued any patent, and 0 otherwise.\\
Consolidated Accounts & A binary variable equal to one if the firm
consolidates accounts of subsidiaries\\
NACE rev. 2 & A 2-digit industry affiliation following the European Classification\\
NUTS 2-digit & The region in which the company is located following the European classification.\\
Productive Capacity & It is an indicator of investment in productive capacity computed as $\frac{Fixed\ Assets_t}{Fixed\ Assets_{t-1}+Depreciation_{t-1}}$\\
Capital Intensity & It is a ratio between fixed assets and number of employees for the choice of factors of production.\\
Labour Productivity & It is a ratio between value added and number of employees for the average productivity of labor services.\\
Interest Coverage Ratio (ICR) &  It is a ratio between EBIT and Interest Expenses, as yet another proxy of financial constraints as in \cite{caballero2008zombie}.\\
TFP & It is the Total Factor Productivity of a firm computed as in \citet{ackerberg2015identification}.\\
Financial Constraints & It is a proxy of financial constraints as in \cite{nickell1999does}, calculated as a ratio between interest payments and cash flow\\
\hline
\end{tabular}
\end{table}
\newpage

\large{Table A1: Panel (B): List of predictors}\\

\begin{table}[!htpb]
\begin{tabular}{p{8cm}p{8cm}}
\hline \hline
\textbf{Variable}& Description\\
\hline

Markup & It an estimate of a firm's markup following \cite{DeLoeckerWarzynski}.\\
ROA & It is a ratio of EBITDA on Total Assets for returns on assets.\\
Financial Sustainability & It is a ratio between Financial Expenses and
Operating Revenues.\\
Size-Age &  It is a synthetic indicator proposed by Hadlock and
Pierce (2010), computed as $(-0.737\cdot log(totalassets) )+(
0.043 \cdot log(totalassets))^2 -( 0.040 \cdot age$ to catch the non-linear relationship between financial constraints, size and age. \\
Capital Adequacy Ratio &  It is a ratio of Shareholders' Funds over Short and
Long Term Debts. \\
Liquidity Ratio & A ratio between Current Assets minus Stocks and Current Liabilities.\\
Liquidity Returns & It is a ratio between Cash Flow and Total Assets\\
Regional Spillovers & It is a proxy proposed by \cite{bernard2004some} computed as a share of exporting plants out of total plants in a region.\\
Industrial spillovers & It is a proxy proposed by \cite{bernard2004some} computed as a share of exporting plants on total plants in a 2-digit industry.\\
External Economies of Scale & It is a proxy proposed by \cite{bernard2004some} computed as a share of exporting plants out of the total in an industry-region cell.\\
Size & Measure of firm size computed as (log of) number of employees.\\
Average Wage Bill & It is computed as ( log of) costs of employees divided by number of employees.\\
Inward FDI & It is a binary variable with value 1 if the firm has foreign headquarters and 0 otherwise.\\
Outward FDI & It is a binary variable with value 1 if the firm has subsidiaries abroad and 0 otherwise.\\
\hline
\end{tabular}
\end{table}

\newpage

\section*{Appendix B: Figures and Tables}
\setcounter{table}{0}
\renewcommand{\thetable}{B\arabic{table}}
\setcounter{figure}{0}
\renewcommand{\thefigure}{B\arabic{figure}}
\centering

\begin{table}[H]
\caption{Sample coverage - size classes}
\label{tab: coverage size_industry}
\centering
\resizebox{0.95\textwidth}{!}{%
\begin{tabular}{lcccccccccccc}
\hline 
\multirow{2}{1.5cm}{NACE rev.2}&\multicolumn{6}{c}{Sample  - N. employees}&\multicolumn{6}{c}{Population - N. employees}\\
        &0-9&10-19&20-49&50-249&250+&Total&0-9&10-19&20-49&50-249&250+&Total\\
        \hline
        10&1,649&711&611&488&172&3,631&45,798&3,225&1,382&679&204&51,288\\
        11&233&105&93&59&21&511&3,397&205&147&76&28&3,853\\
        13&93&76&107&80&7&363&4,586&209&151&113&17&5,076\\
        14&117&51&49&47&22&286&9,391&140&89&57&16&9,694\\
        15&43&24&36&47&16&166&3,038&70&69&45&21&3,243\\
        16&274&182&178&93&8&735&8,869&560&337&168&21&9,956\\
        17&48&64&105&129&39&385&865&123&121&120&62&1,292\\
        18&381&144&167&86&6&784&14,455&445&277&123&17&15,316\\
        19&1&3&4&6&5&19&NA&NA&3&3&7&25\\
        20&134&109&177&223&87&730&NA&NA&190&219&99&2,515\\
        21&16&18&36&58&61&189&NA&NA&31&50&55&252\\
        22&192&173&274&279&53&971&1,963&405&431&319&86&3,205\\
        23&348&135&161&136&59&839&7,094&266&234&136&72&7,803\\
        24&39&33&53&122&51&298&377&60&56&70&35&599\\
        25&988&792&869&571&75&3,295&13,917&2,174&1,498&734&136&18,460\\
        26&134&113&136&154&70&607&1,700&219&157&171&49&2,295\\
        27&106&83&120&123&64&496&1512&169&168&136&63&2,048\\
        28&281&171&320&319&101&1,192&2,983&455&536&399&160&4,534\\
        29&84&62&103&157&98&504&1,092&156&160&152&75&1,635\\
        30&36&22&30&70&41&199&838&57&63&95&55&1,107\\
        31&148&55&78&66&9&356&8,976&164&134&68&13&9,356\\
        32&311&121&108&102&26&668&20,551&394&217&133&44&21,338\\
        \hline
        Total & 5,656 & 3,248 & 3,816 & 1, 091& 3,415 & 17,226& 151,402 & 9,496, & 6,451 & 4,066 & 1,335 & 174,898\\
        \hline
    \end{tabular}}
\begin{tablenotes}
\footnotesize
\singlespacing
\item Note: French manufacturing firms are sourced from Orbis, by Bureau Van Dijk. Sample coverage by number of employees in 2017 (left panel) is compared with information on population sourced from EUROSTAT Structural Business Statistics. Please note that number of employees may report missing values from sample data, thus number of observations do not sum up to sample totals.
\end{tablenotes}
\end{table}

\newpage

\begin{table}[H]
\caption{Prediction accuracies after cross-validating training and testing sets}
    \label{tab:bart_mia_cv}
    \centering
        \resizebox{0.9\textwidth}{!}{%

    \begin{tabular}{lccccc}
    \hline \hline
    Measure &Sample 1&Sample 2&Sample 3&Sample 4&Sample 5\\
    \hline
Sensitivity&0.649&0.647&0.654&0.65&0.648\\
Specificity&0.911&0.904&0.905&0.905&0.907\\Balanced Accuracy&0.780&0.775&0.780&0.778&0.778\\
ROC&0.909&0.903&0.907&0.903&0.908\\
PR&0.739&0.738&0.742&0.732&0.739\\
\hline
N.Obs&103,540&102,748&102,169&102,028&101,712\\
\end{tabular}}
\begin{tablenotes}
\footnotesize
\singlespacing
\item Note: We report prediction accuracies of BART-MIA after cross-validating the algorithm on five different random training and testing sets. Our aim is to check whether predictions are robust against data sampling.
\end{tablenotes}

\end{table}

\begin{table}[H]
\caption{Prediction accuracies with optimal thresholds \citep{liu2012classification}}
    \label{tab: optimal_cutoff}
    \centering
    \begin{tabular}{lccccccc}
    \hline \hline
    Model&Sensitivity&Specificity&Balanced Accuracy&ROC&PR&Threshold\\
    \hline
    Logit-Lasso&0.786&0.676&0.716&0.785&0.789&0.513\\
    Logit&0.760&0.688&0.724&0.794&0.805&0.517\\
    Random forest&0.760&0.686&0.723&0.795&0.801&0.560\\
    BART&0.730&0.708&0.719&0.791&0.800&0.569\\
    BART-MIA&0.863&0.791&0.827&0.905&0.738&0.280\\

\hline

\end{tabular}

\begin{tablenotes}
\footnotesize
\singlespacing
\item Note: We report prediction accuracies when we select the optimal prediction threshold following \cite{liu2012classification}. 
\end{tablenotes}
\end{table}

\begin{table}[H]
\caption{Prediction accuracies with a subset of predictors}
    \label{tab: subset_predictors}
    \centering
    \begin{tabular}{lcccccc}
    \hline \hline
    Model&Sensitivity&Specificity&Balanced Accuracy&ROC&PR\\
    \hline
    Logit-Lasso&0.668&0.768&0.718&0.786&0.785\\
    CART&0.512&0.907&0.710&-&-\\
    Random forest&0.810&0.627&0.719&0.791&0.793\\
    BART&0.807&0.629&0.718&0.790&0.791\\
    BART-MIA&0.623&0.914&0.768&0.902&0.725\\

\hline

\end{tabular}

\begin{tablenotes}
\footnotesize
\singlespacing
\item Note: We report prediction accuracies after reducing the battery of predictors from $52$ to $23$ variables selected by a robust LASSO \citep{robustlasso}. 
\end{tablenotes}
\end{table}

\begin{table}[H]
\caption{Prediction accuracies after training and testing on separate years}
    \label{tab:bart_mia_annual}
    \centering
    \resizebox{0.9\textwidth}{!}{%
    
    \begin{tabular}{lcccccccc}
    \hline \hline
    Measure &2011&2012&2013&2014&2015&2016&2017&2018\\
    \hline
Sensitivity&0.907&0.896&0.885&0.896&0.901&0.918&0.924&0.928\\
Specificity&0.637&0.632&0.641&0.627&0.639&0.651&0.652&0.654\\
Balanced Accuracy&0.772&0.764&0.763&0.761&0.770&0.784&0.788&0.791\\
ROC&0.903&0.889&0.886&0.888&0.894&0.910&0.919&0.930\\
PR&0.759&0.718&0.725&0.723&0.722&0.729&0.734&0.727\\
\hline
N.Obs&11,375&11,377&11,378&11,383&11,386&11,392&11,388&11,387\\
\end{tabular}}
\begin{tablenotes}
\footnotesize
\singlespacing
\item Note: We report prediction accuracies of BART-MIA after training and testing on separate years. Our aim is to check whether predictions are robust along the timeline.
\end{tablenotes}
\end{table}

\begin{table}[H]
\caption{Prediction accuracies of exporters defined \textit{\'a la} \cite{BekesMurakozy}}
    \label{tab: bart_mia_BM}
    \centering
    \resizebox{0.9\textwidth}{!}{%
    \begin{tabular}{lcccccc}
    \hline \hline
    Exporter Class &Sensitivity&Specificity&  Balanced&ROC &PR& Num.\\
     &&& Accuracy&&& Obs.\\[0.5ex]
    \hline

Permanent Exporters &0.723&0.779&0.751&0.849&0.934&76,185\\
Temporary Exporters &0.421&0.820&0.621&0.755&0.447&73,647\\
Non-Exporters &&0.949&&&&158,625\\
\hline
Total & 0.650 & 0.9066 & 0.7783&0.9048&0.7383&232,272
\end{tabular}}
\begin{tablenotes}
\footnotesize
\singlespacing
\item Note: We report prediction accuracies after BART-MIA for firms classified according to \citet{BekesMurakozy}: i) \textit{permanent exporters} are firms that export at least four consecutive years; ii) \textit{temporary exporters} are remaining firms that export at least once; iii) \textit{non-exporters} are firms that never export.
\end{tablenotes}
\end{table}

\begin{table}[H]
\caption{Prediction accuracies after an exporters' definition based on thresholds of the share of export revenues over total revenues}
    \label{tab: bart_mia_percentiles}
    \centering
    \resizebox{0.9\textwidth}{!}{%
    
    \begin{tabular}{lcccc}
    \hline \hline
Measure&$1^{st}$ Percentile&$2^{nd}$ Percentile&$5^{th}$ Percentile&Benchmark\\
\hline
Sensitivity&0.652&0.641&0.625&0.658\\
Specificity&0.835&0.837&0.852&0.833\\
Balanced Accuracy&0.744&0.739&0.738&0.745\\
ROC&0.836&0.835&0.836&0.836\\
PR&0.737&0.731&0.724&0.738\\
\hline
N.Obs&41,911&41,911&41,911&41,911\\
\end{tabular}}
\begin{tablenotes}
\footnotesize
\singlespacing
\item Note: We report prediction accuracies of BART-MIA after defining as exporters the firms with share of export revenues over total revenues above some specific thresholds, at the $1^{st}$,$2^{nd}$, and $5^{th}$ percentiles of the distribution of the share of export revenues over total revenues.
\end{tablenotes}
\end{table}

\begin{table}[H]
    \centering
        \caption{Prediction accuracies - Imputation of missing values}
        \label{tab: imputed_data}
        \resizebox{0.8\textwidth}{!}{%
     \begin{tabular}{lcccccr}
    \hline\hline
    &Specificity&Sensitivity&Balanced&ROC&PR& N. obs.\\
    &&&Accuracy&&&\\
    \hline
         LOGIT&0.817&0.751&0.784&0.784&0.528&382,606\\ LOGIT-LASSO&0.913&0.541&0.727&0.880&0.682&382,606\\
         CART&0.893&0.617&0.755&&&382,606\\
         Random Forest&0.910&0.647&0.778&0.907&0.738&382,606\\
         BART&0.910&0.635&0.772&0.905&0.731&382,606\\
         \hline
    \end{tabular}}
    \begin{tablenotes}
\footnotesize
\singlespacing
\item Note: For a robustness check, we report prediction accuracies after an imputation of missing values based on median values, while adding a predictor indicating the number of missing entries by observation (number of missing values by row).
\end{tablenotes}
\end{table}
}
\newpage
\section*{\centering Appendix C: Evaluation of prediction accuracy}
\label{appendix : c}
\vspace{0.5in}
\setcounter{table}{0}
\renewcommand{\thetable}{B\arabic{table}}
\setcounter{figure}{0}
\renewcommand{\thefigure}{B\arabic{figure}}

Different metrics are used to evaluate prediction accuracy of machine learning algorithms. Briefly, prediction accuracy metrics compare the classes predicted by the algorithm with the actual ones. In the case of a binary outcome, the comparison generates four classes of results:
\begin{itemize}
    \item \textbf{True Positives:} cases when the actual class of the data point is 1 (Positive) and the predicted is also 1 (Positive);
    \item \textbf{False Positives:} cases when the actual class of the data point is 0 (Negative) and the predicted is 1 (Positive);
    \item \textbf{False Negatives: } cases when the actual class of the data point is 1 (Positive) and the predicted is 0 (Negative);
    \item \textbf{True Negatives:} cases when the actual class of the data point is 0 (Negative) and the predicted is also 0 (Negative);
\end{itemize}
In an ideal scenario we want to minimize the number of False Positives and False Negatives.
\begin{table}[H]
\caption{Confusion Matrix}
\label{tab: conf_matrix}
    \setlength{\extrarowheight}{2pt}
    \begin{tabular}{cc|c|c|}
      & \multicolumn{1}{c}{} & \multicolumn{2}{c}{Actual}\\
      & \multicolumn{1}{c}{} & \multicolumn{1}{c}{$Positives\ (1)$}  & \multicolumn{1}{c}{$Negatives\ (0)$} \\\cline{3-4}
      \multirow{2}*{Predicted}  & $Positives\ (1)$ & True Positives (TP)& False Positives (FP)\\\cline{3-4}
      & $Negatives\ (0)$ & False Negatives (FN) & True Negatives (TN) \\\cline{3-4}
    \end{tabular}
  \end{table}
The metrics we use to evaluate prediction accuracy in our exercises are based on the relationship between the sizes of the above classes.
\paragraph{Sensitivity (or Recall) }
Sensitivity (or Recall) is a measure of the proportion of correctly Predicted Positives, out of the total Actual Positives.
\begin{equation*}
    Sensitivity=\frac{True \:Positives}{True\:Positives + False\:Negatives}
\end{equation*}
\paragraph{Specificity}
Specificity is a measure that catches the proportion of correctly Predicted Negatives, out of total Actual Negatives. 
\begin{equation*}
    Specificity=\frac{True\:Negatives}{True\:Negatives + False\:Positives}
\end{equation*}
\paragraph{Balanced Accuracy (BACC)}
The  Balanced Accuracy (BACC) is a combination of Sensitivity and Specificity. It is particularly useful when classes are imbalanced, i.e., when a class appears much more often than the other. It is computed as the average between the rate of True Positives and the rate of True Negatives.
\begin{equation*}
    BACC=\frac{Sensitivity+Specificity}{2}
\end{equation*}
\paragraph{Receiving Operating Characteristics (ROC)} The ROC curve is a graph showing the performance in classification at different thresholds, expressed in terms of the relationship between True Positive Rate (TPR) and False Positive Rate (FPR), defined as follows: \begin{eqnarray*}
True\:Positive\:Rate=\frac{True\: Positives}{True\:Positives + False\:Negatives}
\end{eqnarray*}
\begin{eqnarray*}
False Positive Rate =\frac{False\:Positives}{False\:Positives + True\:Negatives}
\end{eqnarray*}

The Area Under the Curve (AUC) of ROC is then useful to evaluate performance in a bounded range between $0$ and $1$, where $0$ indicates complete misclassification, $0.5$ corresponds to an uninformative classifier, and $1$ indicates perfect prediction.

\paragraph{Precision-Recall (PR)}
 The PR curve is a graph showing the trade-off between Precision and Recall at different thresholds. Note that Precision and Recall are defined as follows:
 
 \begin{eqnarray*}
     Precision=\frac{True\:Positives}{True\:Positives + False\:Positives}
 \end{eqnarray*}
  \begin{eqnarray*}
     Recall=\frac{True\:Positives}{True\:Positives + False\:Negatives}
 \end{eqnarray*}
 
 As for the ROC curve, the PR AUC is used to evaluate the classifier performance. A High AUC represents both high recall and high precision, thus meaning the classifier is returning accurate results (high precision), as well as returning a majority of all the positive results (high recall).
 
\section*{\centering Appendix D: Location Quotients}

\vspace{0.5in}
\setcounter{table}{0}
\renewcommand{\thetable}{B\arabic{table}}
\setcounter{figure}{0}
\renewcommand{\thefigure}{B\arabic{figure}}
  Let us define $\mathcal{I}=\{1,\dots,n\}$ the set of non-exporting  firms and $\mathcal{R}=\{1,\dots,r\}$ the set of regions (NUTS 2-digit). The $r$ partitions of $\mathcal{I}$ by region $j\in \mathcal{R}$ are defined as:
  \begin{equation*}
      I_j \subset \mathcal{I}, j=1,\dots, r\quad s.t. \quad \bigcup_{j=1}^{r} I_j=\mathcal{I}
  \end{equation*} Let $\mathcal{P}$ be the set of non-exporting firms whose exporting score $e$ is above the one of the median firm in the total distribution of non-exporters, i.e.: \begin{equation*}
      \mathcal{P} \subset \mathcal{I} =\{i\in \mathcal{I}: e_i>median(e)\}
  \end{equation*}
  Again we can define the $r$ partitions of $\mathcal{P}$ by region $j\in \mathcal{R}$ as 
  \begin{equation*}
      P_j \subset \mathcal{P}, j=1,\dots, r \quad s.t. \quad \bigcup_{j=1}^{r} P_j=\mathcal{P}
  \end{equation*} 
  The location quotient, for each region $j=1,\dots,r$ is computed as 
  \begin{equation}
      LQ_j=\frac{\#P_j/\#I_j}{\#\mathcal{P}/\#\mathcal{I}}
  \end{equation}
  
In our case, location quotients (LQ) detect concentration of potential exporters in excess of what one would expect from the national distribution. If, for example, region $j$ has $LQ_j=1.5$, it implies that firms with a high trade potential are $1.5$ times more concentrated in such a region than the average.

\end{document}